\documentstyle{elsart}
\input epsf.sty
\newcommand{\mincir}{\raise -2.truept\hbox{\rlap{\hbox{$\sim$}}\raise5.truept
\hbox{$<$}\ }}
\newcommand{\magcir}{\raise -2.truept\hbox{\rlap{\hbox{$\sim$}}\raise5.truept
\hbox{$>$}\ }}
\newcommand{\minmag}{\raise-2.truept\hbox{\rlap{\hbox{$<$}}\raise 6.truept\hbox
{$>$}\ }}
\newcommand{\be}{\begin{equation}}
\newcommand{\ee}{\end{equation}}
\newcommand{\ba}{\begin{eqnarray}}
\newcommand{\ea}{\end{eqnarray}}
\newcommand{\brr}{\begin{array}}
\newcommand{\err}{\end{array}}
\newcommand{\bc}{\begin{center}}
\newcommand{\ec}{\end{center}}

\newcommand{\bq}{\mbox{\bf q}}
\newcommand{\bx}{\mbox{\bf x}}

\newcommand{\lb}{{\left<\right.}}
\newcommand{\rb}{{\left.\right>}}
\newcommand{\hm}{\,h^{-1}{\rm Mpc}}
\newcommand{\vel}{\,{\rm km\,s^{-1}}}

\begin{document}
\begin{frontmatter}
\title{Constraining the Power Spectrum using Clusters}
\author[PG,TS]{S. Borgani},
\author[PD]{L. Moscardini},
\author[AT,TS]{M. Plionis}, 
\author[GREEN,POL1]{K.M. G\'orski},
\author[NMSU]{J. Holtzman},
\author[NMSU]{A. Klypin},
\author[UCSC]{J.R. Primack},
\author[NMSU]{C.C. Smith} and
\author[OX,POL2]{R. Stompor}
\address[PG]{INFN Sezione di Perugia, c/o Dipartimento di 
Fisica dell'Universit\`{a},
via A. Pascoli, I--06100 Perugia, Italy}
\address[TS]{SISSA -- International School for Advanced Studies,
via Beirut 2--4, I--34013 Trieste, Italy}
\address[PD]{Dipartimento di Astronomia, Universit\`a di Padova,
vicolo dell'Osservatorio 5, I--35122 Padova, Italy}
\address[AT]{National Observatory of Athens, Lofos Nimfon, 
Thesio, 18110 Athens, Greece}
\address[GREEN]{Hughes STX Corporation, Laboratory for Astronomy and Solar
Physics, Code 685, NASA/GSFC, Greenbelt, MD 20771, USA}
\address[POL1]{On leave from Warsaw University Observatory, 
Aleje Ujazdowskie 4, 00-478 Warsaw, Poland}
\address[NMSU]{Department of Astronomy, New Mexico 
State University, Las Cruces, NM 88003, USA}
\address[UCSC]{Physics Department, University of California, Santa Cruz, CA
95064, USA}
\address[OX]{University of Oxford, Department of Physics, Keble Rd, Oxford 
OX1 3RH, UK}
\address[POL2]{Copernicus Astronomical Center, Bartycka 18, 00-716 
Warsaw, Poland}

\begin{abstract}
We analyze an extended redshift sample of Abell/ACO clusters and compare
the results with those coming from numerical simulations of the cluster
distribution, based on the truncated Zel'dovich approximation (TZA), for a
list of eleven dark matter (DM) models. For each model we run several
realizations, so that we generate a set of 48 independent mock Abell/ACO
cluster samples per model, on which we estimate cosmic variance effects.
Other than the standard CDM model, we consider {\bf (a)} $\Omega_0=1$ CDM
models based on lowering the Hubble parameter and/or on tilting the
primordial spectrum; {\bf (b)} $\Omega_0=1$ 
Cold+Hot DM models with $0.1\le \Omega_\nu
\le 0.5$; {\bf (c)} low--density flat 
$\Lambda$CDM models with $0.3\le \Omega_0\le
0.5$. We compare real and simulated cluster distributions by analysing
correlation statistics, the probability density function, and supercluster
properties from percolation analysis. We introduce a generalized
definition of the spectrum shape parameter $\Gamma$ in terms of
$\sigma_{25}/\sigma_8$, where $\sigma_r$ is the rms fluctuation amplitude
within a sphere of radius $r$. As a general result, we find that the
distribution of galaxy clusters provides a constraint only on the shape of
the power spectrum, but not on its amplitude: a shape parameter $0.18
\mincir \Gamma \mincir 0.25$ and an effective spectral index at the $20\hm$
scale $-1.1\mincir n_{\rm eff}\mincir -0.9$ are required by the Abell/ACO
data. In order to obtain complementary constraints on the spectrum
amplitude, we consider the cluster abundance as estimated using the
Press--Schechter approach, whose reliability is explicitly tested 
against N--body simulations. By combining results from the analysis of the
distribution and the abundance of clusters we conclude that, of the
cosmological models considered here, the only viable models are either
Cold+Hot DM ones with $0.2\mincir \Omega_\nu \mincir 0.3$, better if
shared between two massive $\nu$ species, and $\Lambda$CDM ones with
$0.3\mincir \Omega_0\mincir 0.5$.
\end{abstract}

\end{frontmatter}

\section{Introduction}
The generally accepted picture for the formation of the large--scale
structure of the Universe is based on the assumption that gravitational
instability acts on the tiny density perturbations that were present at
the end of recombination (at redshift $z\sim
1000$), and this gives rise to the
observed cosmic structures at $z\mincir 5$. This framework, thanks to the
strong support received from the detection of the large--scale cosmic
microwave background (CMB)
temperature anisotropies (e.g., Bennett et al. 1996;
G\'orski et al. 1996), offers a unique
possibility to make testable predictions about the geometry of the
Friedmann background, as well as about the nature and the amount of dark
matter (DM).

The Cold Dark Matter (CDM) scenario (e.g., Blumenthal et al. 1984;
see also Liddle \& Lyth 1993) provided in past
years a powerful interpretative framework for observations of
large--scale cosmic structures and dynamics. This model relies on the
assumption that DM particles became non--relativistic at an early
epoch, when the mass contained within the horizon was much smaller
than the typical galaxy mass scale, $\sim 10^{11}M_\odot$. In its
standard formulation, the CDM picture is based on a critical density
Universe (with a density parameter $\Omega_0=1$),
$h=0.5$ for the Hubble parameter, and primeval
adiabatic fluctuations provided by inflation with random phases and
Harrison--Zel'dovich spectrum, $P(k)\propto k$. However, once normalized
on large--scales to match the {\em COBE} observations, the CDM power
spectrum turns out to have too much power on small scales ($\mincir
10\hm$) to produce the observed number density of galaxy clusters (e.g., White,
Efstathiou \& Frenk 1993). In addition, its shape at intermediate scales
($20\mincir R \mincir 50 \hm$) is strongly discrepant with respect to
measurements of clustering in the galaxy (e.g., Maddox et al. 1990; Loveday
et al. 1992; Park et al. 1994) and in the
cluster (e.g., White et al. 1987; Bahcall \& Cen 1992; Olivier et al. 1993;
Dalton et al. 1994; Plionis et al. 1995) distributions.

Despite all these failures, the standard Cold Dark Matter (SCDM) scenario
remained a sort of reference model, most of the currently viable
alternative models being elaborated as modifications of SCDM.
Remaining within the random--phase hypothesis, all such alternatives
go in the direction of modifying the power spectrum shape
and/or dark matter composition, so as to
decrease the small--to--large scale power ratio.

A first possibility is adding a ``hot" DM component which is provided by
light neutrinos with mass $m_\nu \sim 5$ eV. The resulting Cold+Hot
Dark Matter (CHDM) scenario is characterized by the suppression of the
small--scale fluctuation growth rate, due to the free streaming of the
light neutrinos. Preliminary computations for this model date back to
the early '80s (e.g., Bonometto \& Valdarnini 1984; Fang, Li \& Xiang 1984;
Achilli, Occhionero \& Scaramella 1985). Subsequent more refined investigations
based both on linear theory (e.g., Holtzman 1989; van Dalen \& Schaefer
1992; Holtzman \& Primack 1993; Pogosyan \& Starobinsky 1995; Liddle et al.
1995a)
and N--body simulations
(e.g., Davis, Summers \& Schlegel 1992; Klypin et al. 1993; Jing \& Fang 1994;
Ma \& Bertschinger 1994; Ghigna et al. 1994; Klypin et al. 1995)
indicate that a hot component fraction $\Omega_\nu \simeq 0.2$ is required
by a variety of observational constraints. A virtue of the CHDM scenario
lies in the fact that neutrinos are known to exist, and that
its validity can be confirmed or disproved
in the next few years thanks to the ongoing experiments
(e.g. Athanassopoulos et al. 1995, and refs. therein)
for the measurement of a possible non--vanishing neutrino mass in the
range of cosmological interest ($m_\nu \magcir 1$ eV). Indeed, standard
arguments predict that $\Omega_\nu \simeq (m_\nu / 92\,{\rm eV})\,
h^{-2}$ is the contribution to the density parameter from one species
$i$ of massive $\nu_i + \bar\nu_i$.
Note also that the possible detection of $m_\nu \sim 5$
eV would be indirect evidence for the existence of CDM, since a
low--density hot Universe would be even more in trouble than the usual
$\Omega_\nu=1$ Hot Dark Matter scenario (Primack et al. 1995).

A further possibility to modify the SCDM is by ``tilting" the primordial
spectrum shape to $P(k) \propto k^n$, with $n<1$, while maintaining the
shape of the pure CDM transfer function. The resulting spectrum turns out
to be ``redder" (i.e., have less short-wavelength power)
than that of the SCDM, as required by the observational
constraints. The possibility of obtaining departures from the
Harrison--Zel'dovich spectrum in the framework of inflationary models has
been investigated for several years (e.g., Lucchin \& Matarrese
1985; Hodges et al. 1990; Adams et al. 1993; Turner, White \& Lidsey 1993).
Comparisons between model predictions and observations (e.g., Cen et al. 1992;
Tormen et al. 1993; Liddle \& Lyth 1993; Moscardini et al. 1995;
White et al. 1995)
indicate that $0.8\mincir n\mincir 0.9$ as a compromise between the need
to substantially change the power spectrum shape and the requirement not
to underproduce large--scale peculiar velocities. Note that
predictions of the tilted CDM models also depend on the ratio between
scalar (density) and tensor (gravitational wave) fluctuation modes, which
may (but need not) be produced by inflationary scenarios providing $n<1$.
As a result, the spectrum normalization can be decreased with respect to the
case of vanishing tensor--mode contribution by an amount which depends
on the assumed inflation model.  General arguments have recently been
given (Ross \& Sarkar 1996) that supersymmetric cosmological models will
have negligible tensor-mode fluctuations.

Another possibility to improve the SCDM is by increasing the size of
the horizon at the matter--radiation equality epoch $t_{eq}$.
The CDM power spectrum
displays at this scale a smooth transition from the large--scale
behaviour, where the primordial spectrum is preserved, and the
small--scale bending of $P(k)$ due to the stagnation of growth for
fluctuations crossing the horizon before $t_{eq}$. Since this
scale is proportional to $(\Omega_oh)^{-1}$, it can
be increased by decreasing either $h$ or $\Omega_0$. While it is not
clear whether current estimates of the Hubble parameter allow for
values substantially smaller than $h=0.5$ (e.g., Branch, Nugent \& Fisher
1995; Sandage et al. 1996; see also 
Mould et al. 1995, and Kennicutt, Freedman \& Mould 1995
for recent reviews on $H_0$ determinations), values of
$\Omega_0 <1$ are surely viable on observational grounds. The presence of
a suitable cosmological constant term
$\Omega_\Lambda \equiv \Lambda/(3H_0^2)$, such that
$\Omega_0+\Omega_\Lambda=1$, can be invoked to account for the
inflationary preference for an almost flat spatial geometry. Note that,
while the presence of the cosmological constant term does not affect the
shape of the transfer function very much,
it changes the resulting CMB temperature
fluctuations and, therefore, the large--scale normalization (e.g.,
G\'orski et al. 1995; Stompor, G\'orski \& Banday 1995). The value of
$\Omega_0$ or, more precisely, of $\Omega_0h$ has to be chosen on the
basis of observational constraints; in particular, the observed
clustering of galaxies is rather well reproduced by a CDM power spectrum
profile with $0.2\mincir \Omega_0h\mincir 0.3$ (see, e.g., Peacock \&
Dodds 1994).

It is clear that this large variety of DM models has a twofold
implication as far as the search for the ``best model'' is
concerned. On the one hand, one has to choose an observable quantity
to be compared with the model predictions which is both robust and as
discriminative as possible among the different models. On the other
hand, the methodology on which this comparison is based must be
flexible and easy to implement, so that many DM models can be tested
at once.

In this respect, the study of the distribution of galaxy clusters
represents a powerful tool to analyze the structure of the Universe on
large scales and to compare it with model predictions. Already
from the first investigations of the correlation properties of Abell
(1958) clusters (e.g., Bahcall \& Soneira 1983; Klypin \& Kopilov 1983;
see Bahcall 1988 for a review) it is has been realized that their
clustering is significantly amplified with respect to that of galaxies, and
thus it allows the detection of a reliable correlation signal on large scales,
where galaxy correlations fade away. For these reasons observational
efforts have extended the Abell
sample to the southern hemisphere (Abell, Corwin \& Olowin 1989) and
recently provided new catalogues based on automated and objective cluster
identification procedures, based both on optical (Dalton et al. 1994;
Collins et al. 1995) and X--ray
(Nichol, Briel \& Henry 1994; Romer et al. 1994) observations.

The classical result for the two--point
cluster correlation function is that it is well modelled by the power--law
\be
\xi(r)\,=\,\left({r_0\over r}\right)^\gamma
\label{eq:xicl}
\ee
on the scale range $10\mincir r\mincir 50 \hm$, with $\gamma\simeq
1.6$--2 and $r_0\simeq 16$--25$\hm$, depending both on the cluster
richness and on the considered observational sample (see, e.g., Postman,
Huchra \& Geller 1992; Peacock \& West 1992; Nichol et al. 1992; Dalton et
al. 1994). On the other hand, the reliability for
the power--law model of $\xi(r)$ has been questioned by Olivier
et al. (1993). Note that the large amplitude of the cluster
correlation represented one of the first pre--{\em COBE} failures of the
SCDM model. Indeed, White et al. (1987) showed that this model produces
a too small $\xi(r)$, with $r_0\simeq 10\hm$, a value that cannot be
reconciled with observational data, even by invoking a substantial
overestimate of cluster correlations, due to richness contamination from
projection effects (e.g., Sutherland et al. 1988; see, however,
Jing, Plionis \& Valdarnini 1992).  Olivier et al. (1993) emphasized
that the problem with SCDM is that it predicts that $\xi(r)$ becomes
negative for $r\magcir 30 \hm$, while observationally $\xi(30 \hm)$
is clearly positive.

A further advantage in using clusters to trace the distribution of DM
lies in the fact that their typical mass scale ($\sim 10^{15}M_\odot$) is still
in the quasi--linear or mildly non--linear regime of gravitational
clustering. Therefore, cluster positions should be correctly predicted
without resorting to the whole non--linear dynamical description provided
by N--body codes (e.g., Bahcall \& Cen 1992; Croft \& Efstathiou 1994;
Klypin \& Rhee 1994). On the other hand, analytical approaches based both
on linear theory (e.g., Lumsden, Heavens \& Peacock 1989; Coles 1989;
Borgani 1990; Holtzman \& Primack 1993) and on the Zel'dovich (1970)
approximation (Doroshkevich \& Shandarin 1978; Mann,
Heavens \& Peacock 1993) can only deal with low--order correlations,
while a more complete statistical description and the inclusion of
observational biases and error analysis can be hardly approached.

A useful alternative to the exact N-body solution lies in the implementation
of numerical simulations
in which the gravitational dynamics is described by the Zel'dovich
approximation (ZA hereafter; Blumenthal, Dekel
\& Primack 1988; Borgani, Coles \& Moscardini 1994). The ZA has been
shown to provide an excellent description of the evolution of the density
field, at least on scales where $\sigma \mincir 1$ for the rms
fluctuation amplitude, once the initial (linear) density fluctuations are
smoothed on a suitably chosen scale (e.g., Coles, Melott \& Shandarin
1993; Melott, Pellman \& Shandarin 1994; Sathyaprakash et al. 1995).
Borgani et al. (1995; BPCM hereafter) showed in detail that
this ``truncated"
version of the ZA (TZA hereafter) provides a fully reliable description
of the cluster distribution, when compared with the outputs of analogous
N--body simulations. The remarkable
advantage of using the TZA cluster simulations with respect to N--body
ones lies in the fact that the former are realized with a single
time--step, thus requiring a tiny fraction of the computational cost of
the latter. Of
course, the price to be paid with respect to the N--body treatment is
that, while individual cluster positions are well reproduced, the cluster
masses are not. 
Indeed, the TZA fails to account for those aspects of non--linear 
dynamics, like merging of surrounding structures and virialization,
which determine the mass of a cluster.
For this reason, in the following we will base our
discussion about the cluster mass function not on the TZA simulations,
but on the analytical approach provided by the Press \& Schechter (1974)
theory. We will also present a quantitative comparison of 
N--body cluster simulations and PS predictions, that confirms the 
reliability of this approach (cf. also White et al. 1993; Lacey \& Cole 1994;
Mo, Jing \& White 1996).

In this paper we take advantage of the reliability of the TZA to perform
cluster simulations for a list of eleven DM models, belonging to the
above described categories. For each model we run several realizations
and extract from each of them mock samples, which reproduce the same
observational characteristics of a combined Abell/ACO cluster redshift
catalogue. We provide a set of 48 independent samples for
each considered model to which we apply the same statistical analyses as
for the observational one.

The plan of the paper is as follows. In Section 2 we present the
observational redshift sample of Abell/ACO clusters. In Section 3 we
describe the simulations and the considered DM models. Section 4 is
devoted to the description of the statistical analyses applied to both
observational and simulated cluster samples and to the presentation of
the results:
they concern the correlation statistics, the probability
density function (PDF) and supercluster properties based on
percolation analysis.
In Section 5 we show how the distributions of
cluster positions and cluster masses convey complementary information
about the shape and the
amplitude of the model power spectrum, respectively. In Section 6 we
draw our main conclusions.

\section{The Abell--ACO cluster sample}
\subsection{Sample definition}
We use an updated version of the combined Abell/ACO (Abell 1958; Abell, Corwin
\& Olowin 1989) $R\ge 0$ cluster sample, as was initially defined and analysed
in Plionis \& Valdarnini (1991, hereafter PV; 1995)  and in BPCM.
We have included the new redshifts provided by the ESO Abell cluster survey
(Katgert et al. 1996) and by Quintana \& Ramirez (1995). The northern (Abell
1958) sample, with a declination
$\delta \ge -17^{\circ}$, is defined by those clusters that have
measured redshifts $z\mincir 0.1$,
while the southern sample (ACO; Abell, Corwin
\& Olowin 1989), with $\delta < -17^{\circ}$, is defined by those clusters with
$m_{10}< 17$, where $m_{10}$ is the magnitude of the tenth brightest cluster
galaxy in the magnitude system corrected according to PV. Furthermore we
limit our sample to a maximum distance of $240\hm$ and to
avoid the gross effects of Galactic absorption we only use clusters with
$|b|\ge 30^{\circ}$.

There are in total 409 Abell/ACO clusters fulfilling the above criteria;
255 Abell clusters all having measured redshifts, and 154 ACO clusters, of
which 138 have measured redshifts and the remaining
16 have redshifts $z$ estimated from the $m_{10}$--$z$ relation derived in PV
(cf. also Kerscher et al. 1996). Since we will compare the
real data with simulations based both on flat  cosmological models,
with and without a cosmological constant, we convert redshifts into cluster
distances using the general formula for the distance by apparent
size (cf. Peebles 1993):
\begin{equation}
r(z)~=~{c\over H_0 (1-\Omega_0-\Omega_\Lambda)}\,
\sinh\left({\int_0^z dz'/E(z') \over 1-\Omega_0-\Omega_\Lambda}\right)\,,
\label{eq:dz}
\end{equation}
where $c$ is the speed of light and
\begin{equation}
E(z)=\sqrt{\Omega_0(1+z)^3+\Omega_\Lambda+(1-\Omega_0-
\Omega_\Lambda)(1+z)^2} \nonumber
\end{equation}
with $H_0 = 100 \; h \;$ km sec$^{-1}$ Mpc$^{-1}$. Final results
are actually quite insensitive
to the choice of
the cosmological parameters to be inserted in eq.(\ref{eq:dz}) so,
except where differently specified, we will from now on use distances
based on $(\Omega_0,\Omega_\Lambda)=(0.4,0)$,
which are intermediate between the $(\Omega_0,\Omega_\Lambda)=(1,0)$
and $(\Omega_0,\Omega_\Lambda)=(0.2,0.8)$ cases.

\subsection{Cluster selection functions}
In our analysis we model the effects of Galactic absorption by the
usual cosecant law:
\begin{equation}\label{eq:obs1}
P(|b|) = \mbox{dex} \; \left[ \alpha \left(1 - \csc |b| \right) \right]
\end{equation}
with $\alpha \approx 0.3$ for the Abell sample (Bahcall \& Soneira 1983;
Postman et al. 1989) and $\alpha \approx 0.2$ for the ACO sample
(Batuski et al. 1989). The cluster--redshift selection function, $P(z)$, is
determined in the usual way (cf. Postman et al. 1989), by fitting
the cluster density as a function of $z$:
\begin{eqnarray}
P(z)~=~\left\{ \begin{array}{ll}
1 & \mbox{if $z\le z_c$} \\
A\,\exp{(-z/z_o)} & \mbox{if $z >  z_c$}
\end{array}
\right.
\label{eq:pz}
\end{eqnarray}
where $A=\exp{(z_c/z_o)}$ and $z_c$ is the redshift out to which the space
density of clusters remains constant (volume--limited regime). Using a
best--fit procedure, we obtain $z_c\approx 0.078$, $z_o\approx 0.012$ and
$z_c\approx 0.068$, $z_o\approx 0.014$  for Abell and ACO samples,
respectively. The exponential decrease of $P(z)$ could introduce large
shot--noise
errors when correcting the cluster density at large distances and this is
why we prefer to limit our analysis to $r_{\rm max}=240$ $h^{-1}$ Mpc.

Within the volume--limited regime and correcting for Galactic absorption via
eq.(\ref{eq:obs1}), we obtain  $\langle n\rangle_{\rm Abell}\simeq 1.6 \times
10^{-5} \;(h^{-1} {\rm Mpc})^{-3}$ and $\langle n\rangle_{\rm ACO}\simeq
2.6\times 10^{-5} \;(h^{-1} {\rm Mpc})^{-3}$, for the Abell and ACO cluster
number density respectively. This density difference is mostly spurious, due
to the higher sensitivity of the IIIa--J emulsion plates on which the ACO
survey is based (for more details see Batuski et al. 1989; Scaramella et al.
1990; PV) but part of it is intrinsic, due to the presence of the Shapley
concentration in the ACO sample (Shapley 1930; Scaramella et al. 1989;
Raychaudhury 1989); excluding the small $b>30^{\circ}$ region of the ACO sample
(corresponding to a solid angle $\delta\Omega \mincir 0.08\pi$),
where most of the Shapley concentration lies,
we obtain $\langle n\rangle_{\rm ACO}\simeq 2.3\times 10^{-5}(h^{-1} \; {\rm
Mpc})^{-3}$. Note that the uncertainty of $\langle n\rangle$, estimated by its
fluctuations in equal volume shells of $\delta V \simeq 2\times 10^6 \;
{\rm Mpc}^3$, is $\sim 2.5\times 10^{-6}$.
The above density values correspond to average $R \geq 0$
cluster separations of
$\langle r_{\rm Abell}\rangle \simeq 40 \; h^{-1} {\rm Mpc}$ and
$\langle r_{\rm ACO}\rangle \simeq 35 \; h^{-1} {\rm Mpc}$.

Since we use the combined Abell/ACO sample to derive the various statistical
quantities we need to account for the systematic difference in the 
Abell and ACO
cluster number densities. We do so by using a distance dependent weighting
scheme in which each Abell cluster is weighted according to a radial
function,
\begin{equation}\label{eq:weight}
w(r)=\frac{n_{\rm ACO}(r)}{n_{\rm Abell}(r)}\;\;\,
\end{equation}
We have verified that our final results are quite insensitive 
to whether we weight Abell or ACO clusters with $w$ or $1/w$, respectively.

\section{The cluster simulations}
\subsection{The simulation method}
We follow the evolution of the gravitational clustering on large
scales by resorting to the Eulerian--to--Lagrangian coordinate relation
provided by the ZA (Zel'dovich 1970; Shandarin \&
Zel'dovich 1989), $\bx(\bq,t)=\bq-b(t){\bf \nabla}_{\bq}
\psi( \bq)$. Here $\bx$ and $\bq$ are the initial and the final particle
positions, respectively, $b(t)$ is the fluctuation linear growing mode and
$\psi(\bq)$ the initial gravitational potential. The details and the
advantages of using this approach for simulating the cluster distribution
have been discussed by BPCM (see
also Tini Brunozzi et al. 1995, and Moscardini et al. 1996).

The simulation procedure can be sketched as follows.
\begin{description}
\item[(a)] The linear power spectrum $P(k)$ is filtered with a Gaussian
window, $P(k)\to P(k)e^{-k^2R_f^2}$, so as to suppress the amount of shell
crossing at small scales. Consequently the performance of this truncated
Zel'dovich approximation (TZA) are considerably improved (Coles, Melott \&
Shandarin 1993; Melott, Pellman \& Shandarin 1994). The filtering scale $R_f$
is optimally chosen for each model so as to give $N_s=1.1$ for the number of
streams at each Eulerian point (Kofman et al. 1994; BPCM).
\item[(b)] A random--phase realization of $\psi(\bq)$, based on a given $P( k)$
model, is realized on the grid of the simulation box.
\item[(c)] Particles, initially located at the grid positions, are then moved
according to the TZA.
\item[(d)] The density field is reconstructed on the grid through a TSC
interpolating spline (see e.g. Hockney \& Eastwood 1981) from the particle
positions.
\item[(e)] Clusters, having mean separation $d_{cl}$, are selected as the
$N_{cl}$ highest peaks, $N_{cl}=(L/d_{cl})^3$ being the total number of
clusters expected within a simulation box of size $L$.
\end{description}
BPCM discussed in detail the reliability of the TZA simulation method, by
comparing it with full N--body simulations. In particular,
in that paper three main points were emphasized.
\begin{itemize}
\item TZA simulations reproduce N--body results with great accuracy, as
long as the cluster mass scale is still in the quasi--linear regime,
$\sigma_8\mincir 1$
[$\sigma_{8}$ is the rms fluctuation amplitude within a top--hat sphere of
radius $8\hm$ radius]; the cluster clustering is reproduced not only in a
statistical sense, but even point--by--point.
\item As the cluster mass scale undergoes substantial non--linear
evolution ($\sigma_8\magcir 1$), cluster positions start being affected
by non--linear effects, like merging and infall, which are not accounted
for by the TZA representation. As a result, the cluster clustering remains
remarkably stable in N--body simulations 
(see Fig. 2 in Borgani et al. 1995; cf.\
also Croft \& Efstathiou 1994), while it keeps growing in TZA simulations.
\item Taking advantage of this clustering stability, the TZA simulations can
be performed reliably also to models which require $\sigma_8>1$ by doing
them at a reasonably less evolved stage, when $\sigma_8\mincir 1$.
\end{itemize}

For each of the models that we will consider, we run a set of six
realizations, each within a box of size $L=640\hm$ and using $256^3$ grid
points and as many particles.

\subsection{The models}
We have analysed a list of eleven models, which can be divided into three
main categories.
\begin{description}
\item[(a)] CDM models with $\Omega_0=1$. Within this class of models, we
consider the standard CDM (SCDM) one with $h=0.5$ and scale--free primordial
spectrum, and also
CDM with low Hubble parameter, $h=0.4$, both with $n=1$ (LOWH)
and with $n=0.9$ (LOWH$_{0.9}$) for the primordial spectral index.
\item[(b)] Cold+Hot DM models with different amounts for the hot component,
$\Omega_\nu=0.1,0.2,0.3$ and 0.5, provided by one massive neutrino species
(C$\nu$DM). As a further model, we consider also the $\Omega_\nu=0.2$ case,
with two equally massive neutrinos (C$\nu^2$DM; see Primack et al. 1995).
\item[(c)] Low--density CDM models, with flatness provided by the cosmological
constant term ($\Lambda$CDM), with $\Omega_0=0.3,0.4$ and 0.5 (see Klypin,
Primack \& Holtzman 1996, for a detailed analysis of the $\Lambda$CDM$_{0.3}$
model).
\end{description}

\begin{figure}
\mbox{\epsfxsize=14 cm\epsffile{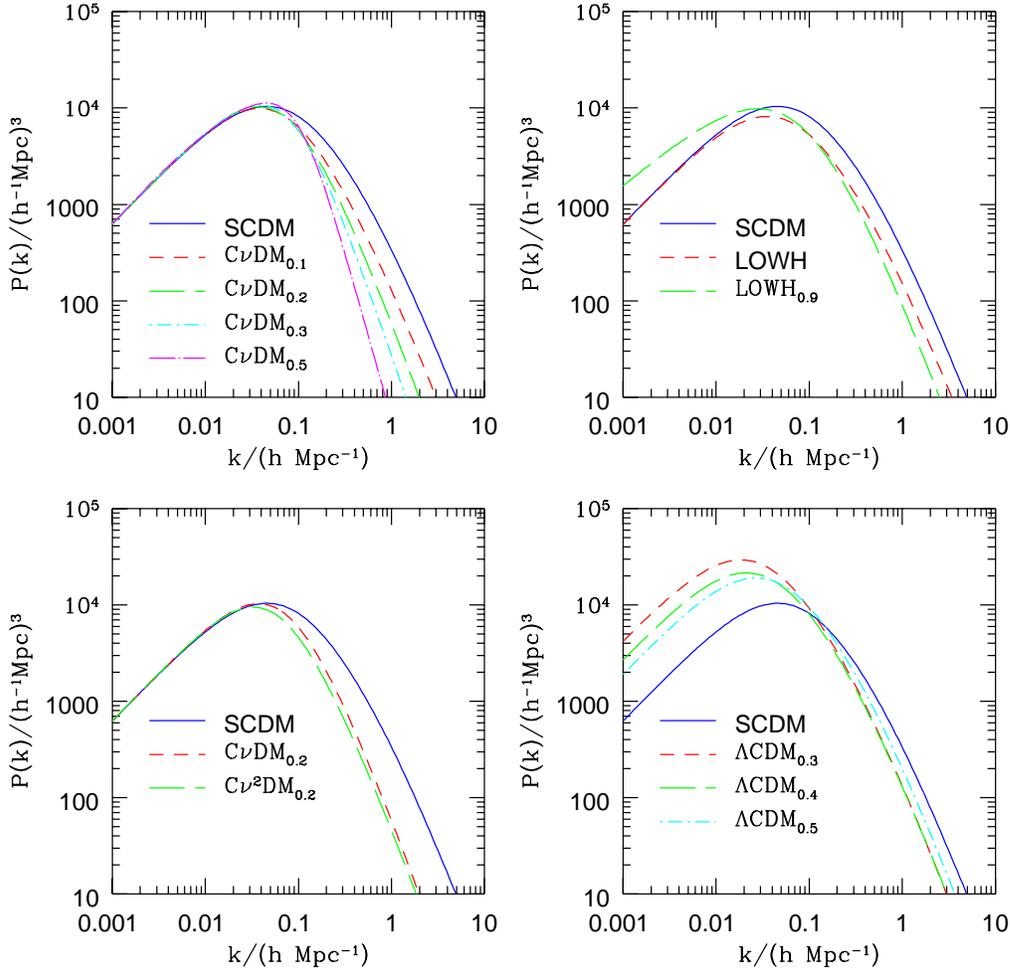}}
\caption{Power spectra for the eleven models considered in this paper.
All models are normalized to the four-year {\em COBE} data.  See Table
1 for details of the models and their normalization.}
\label{fi:powerspectra}
\end{figure}

The parameters defining all eleven models are listed in Table
\ref{t:mod}.  The reported values of $\sigma_8$ in column 8 are from
directly fitting the CMB anisotropies for each model, obtained from a
full Boltzmann code (see Stompor 1994), 
to the four-year {\em COBE} data; the appropriate
$Q_{\rm rms-PS}$ for each case is also given.  These were all fitted
to the data in the same way, but it should be emphasized that there is
about $\pm3\%$ uncertainty in these central values and about $\pm 9\%$
($1\sigma$) overall uncertainty (G\'orski et al. 1996).  The tilted
model with $n=0.9$ (LOWH$_{0.9}$) is normalized assuming zero
contribution from gravitational waves.  

The corresponding {\em COBE}--normalized power spectra for each of
these eleven models are plotted in Figure \ref{fi:powerspectra}.
Several features are immediately apparent.  The amount of small--scale
power is decreased as $\Omega_\nu$ increases in CHDM models, or as $h$
is lowered in CDM models.  The C$\nu^2$DM model has less power on the
cluster scale than the C$\nu$DM model with the same $\Omega_\nu=0.2$.
The $\Lambda$CDM models have somewhat less small-scale power than the
SCDM model, but also much more large--scale power.

Table \ref{t:mod} also has a second $\sigma_8$ column, which gives the
values actually used for the TZA simulations.  As discussed in the
previous subsection, for models with {\em COBE} $\sigma_8>1$, we use a
smaller value in the TZA simulations to avoid excessive non--linear
effects.  For all the $\Omega=1$ models except SCDM, this $\sigma_8$
corresponds to $a_9=7.0\,\mu$K, which is consistent with the one--year 
{\em COBE} normalization (Smoot et al. 1992). As for the $\Lambda$CDM models,
$\sigma_8$ was simply set equal to 0.80 for the TZA calculations.
This procedure is not expected to
represent a limitation, thanks to the above--mentioned stability of the
cluster clustering against changes in the spectrum amplitude.  In
the discussion of the cluster abundance in \S 5, we use the four-year
{\em COBE} normalization given in Table \ref{t:mod}.

\begin{table*}[table]
\centering
\caption[]{The DM models.
Column 1: the name of each model considered in this paper.
Column 2: the density parameter $\Omega_0$.
Column 3: the percentage of the hot component $\Omega_\nu$.
Column 4: the percentage of baryons $\Omega_{bar}$.
Column 5: the Hubble parameter $h$ (in units of 100$\vel$Mpc$^{-1}$).
Column 6: the age of the Universe $t_0$.
Column 7: the rms Quadrupole amplitude for the power spectrum considered.
Column 8: the corresponding rms fluctuation amplitude within a
top--hat sphere of $8\hm$ radius $\sigma_8$.
Column 9: the actual value of $\sigma_8$ used in the TZA simulations.}
\tabcolsep 7pt
\begin{tabular}{lcccccccc} \\ \hline \hline
Model & $\Omega_0$ & $\Omega_\nu$ & $\Omega_{\rm bar}$ & $h$ & $t_0$ &
$Q_{\rm rms-PS}$ & $\sigma_8$ & $\sigma_8$
\\    &            & \%           & \%                 &     & Gyrs  &
&     &  TZA \\
\hline \\
\multicolumn{7}{l}{CDM models} \\
SCDM & 1.0 & 0 & 5.0 & 0.5 & 13.0 &17.5 & 1.19 & 1.14 \\
LOWH & 1.0 & 0 & 7.8& 0.4 & 16.3 & 17.5 & 0.89 & 0.84 \\
LOWH$_{0.9}$ & 1.0 & 0 & 7.8& 0.4 & 16.3 & 19.0 & 0.72 & 0.68 \\ \\
\multicolumn{7}{l}{CHDM models} \\
C$\nu$DM$_{0.1}$ & 1.0 & 10 & 5.0 & 0.5 & 13.0 & 17.5 & 0.92 & 0.89 \\
C$\nu$DM$_{0.2}$ & 1.0 & 20 & 5.0 & 0.5 & 13.0 & 17.5 & 0.80 & 0.77 \\
C$\nu^2$DM$_{0.2}$ & 1.0 & 20 & 5.0 & 0.5 & 13.0 & 17.5 & 0.72 & 0.69 \\
C$\nu$DM$_{0.3}$ & 1.0 & 30 & 5.0 & 0.5 & 13.0 & 17.5 & 0.74 & 0.71 \\
C$\nu$DM$_{0.5}$ & 1.0 & 50 & 5.0 & 0.5 & 13.0 & 17.5 & 0.69 & 0.66 \\ \\
\multicolumn{7}{l}{$\Lambda$CDM models} \\
$\Lambda$CDM$_{0.3}$ & 0.3 & 0 & 2.6 & 0.7 & 13.5 & 20.6 & 1.06 & 0.80 \\
$\Lambda$CDM$_{0.4}$ & 0.4 & 0 & 3.5 & 0.6 & 14.5 & 19.2 & 1.00 & 0.80 \\
$\Lambda$CDM$_{0.5}$ & 0.5 & 0 & 3.5 & 0.6 & 13.5 & 18.4 & 1.13 & 0.80 \\ 
\hline
\end{tabular}
\label{t:mod}
\end{table*}

\subsection{Extraction of mock samples}
We generate a parent population of clusters within the
simulation box, by using $d_{cl}=35\hm$ for their average separation
which corresponds to that of the ACO cluster sample.
We therefore obtain, within each simulation box, the
positions and peculiar velocities of 6114 clusters.

Since the limiting depth of the Abell/ACO sample is much smaller than our
simulation box size, we can identify more than one independent (or nearly
independent) observer, in each simulation. In fact, we maximize the number of
such observers to eight by placing them along the main diagonal
axes of the cube. Each observer is located at a distance of $160\hm$ from
the boundary of the box (i.e., at a distance of $\simeq 277\hm$ from the
nearest corner). The periodic boundary conditions in the $640 \hm$ simulation 
box allow us to periodically replicate this box as required so that each 
observer can define the cluster sample up to the same depth,
$240\hm$, of the observational catalogue. In order to minimize the
overlapping between mock samples extracted in the same box, 
the coordinate system for each observer is such that its
``galactic" plane is defined to be orthogonal to the one of the 
other three nearest
observers. Since we exclude the portion of the sky with $|b|\le
30^\circ$, there is only a modest overlap between neighbouring samples.
We find that typically only 5\% of the clusters are selected twice
in each box. Finally, once the position of an observer is fixed,
velocities of the clusters are used to convert the surrounding cluster
distribution to redshift space.

\begin{figure}
\mbox{\epsfxsize=14 cm\epsffile{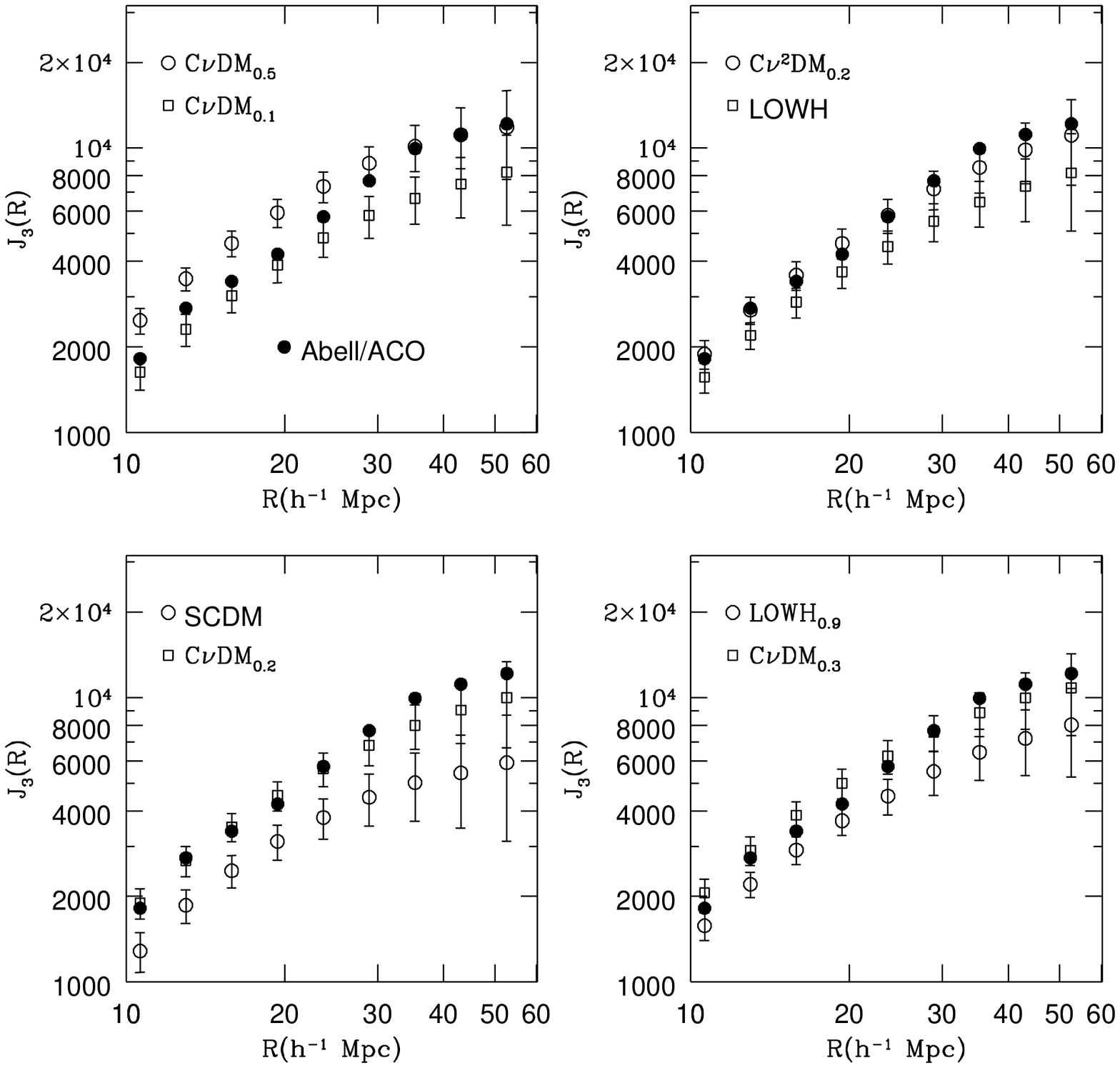}}
\caption{The $J_3(R)$ integral for the Abell/ACO sample (filled circles)
and for $\Omega_0=1$ simulated models (open symbols). Errorbars for simulations
correspond to the $1\sigma$ scatter over the ensemble of 48 observers
(see text).}
\label{fi:j3_chdm}
\end{figure}

Since we have run six realizations for each model
with eight observers per realization, we end up with 48
observers, which sample almost independent volumes. This is a rather large
ensemble on which we can estimate the effects of cosmic variance. Once the
geometry of
the sample boundary has been fixed, the membership of clusters within each
of the samples is decided on the basis of a Monte Carlo rejection method,
which reproduces the same selection effects [i.e., the Galactic absorbtion
of eq.(\ref{eq:obs1}) and the redshift extinction of eq.(\ref{eq:pz})] as
in the real sample. Since the cluster number density is fixed in the whole
simulation box, variations in the number of cluster members in each sample
are expected. We do not attempt to reproduce in each mock sample the exact
number of clusters as in the real one. However, we distinguish the Abell
and ACO portions of the sample volume and degrade randomly the number of
Abell--like clusters, in order to reproduce the same relative number density
variations between the Abell and ACO cluster samples [cf.
eq.(\ref{eq:weight})].

Furthermore, we generate for each observer 10 mock samples, based on
different random realizations of the same selection functions.
This allows us to check whether variations of our results
due to this effect are significant or not. We have verified that
they are always smaller (by a factor $\magcir 2.5$) than the
observer--to--observer (i.e., cosmic variance) scatter. Note that the final
results for each observer are taken as the average over these 10 different
realizations of the selection functions.

\section{Analysis and results}
In this section we compare the results of the statistical analyses that we
applied to both Abell/ACO and simulated data sets. We will deal with
correlation statistics, probability density functions (PDF) and percolation
analysis.

\subsection{The correlation analysis}
In order to provide a robust analysis of the cluster correlation properties,
we estimate the quantity
\be
J_3(R)\,=\,{1\over 4\pi}\int_0^R\xi(r)\,d^3r\,,
\label{eq:j3}
\ee
which is the integral of the two--point correlation function, $\xi(
r)$, within a sphere of radius $R$. According to eq.(\ref{eq:j3}), if we
model the two--point correlation function with the power--law
$\xi(r)\propto r^{-\gamma}$, then $J_3(R)\propto R^{3-\gamma}$.
Therefore, $J_3(R)$ is expected to grow as long as $\xi(r)>0$, while its
flattening is the signature for a rapid declining of the two--point
function. BPCM showed that the integral nature of $J_3(R)$ makes it a
reliable and stable clustering diagnostic. We estimate
$J_3(R)$ by cross--correlating the cluster positions of real and simulated
samples with those of random samples using the estimator:
\be
J_3(R)\,=\,{R^3\over 3}\left(\frac{DD \cdot N_{R}^2}{RR \cdot
N_{D}^2}-1\right)\,,
\label{eq:j3es}
\ee
where $DD$ and $RR$ are the number of data--data and random--random pairs,
respectively, having separation $\le R$ and $N_R$ and $N_D$ are the numbers
of the random and real clusters, respectively. We compute the value
of $RR$ by using 200 random samples having the same selection
functions as the reference real sample and $N_R=3000$.

\begin{figure}
\mbox{\epsfxsize=14 cm\epsffile{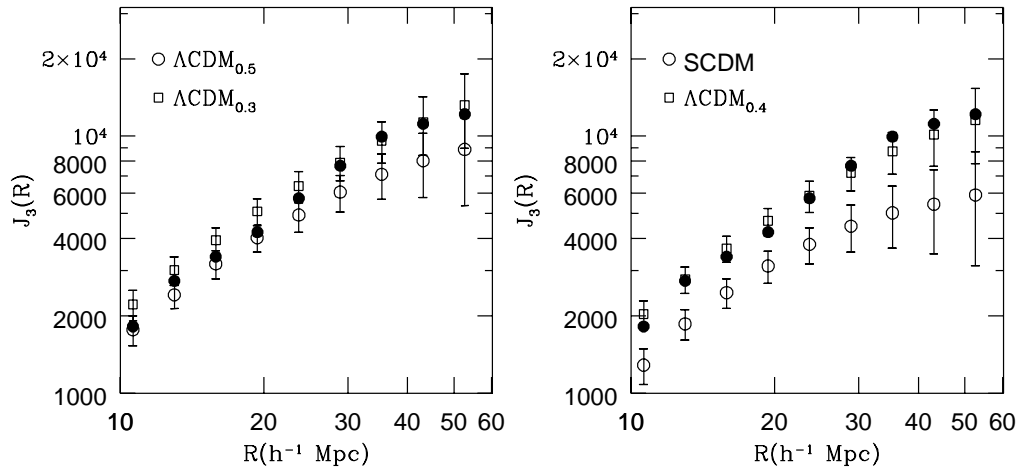}}
\vspace{-8truecm}
\caption{The same as in the previous figure, but for $\Lambda$CDM
models. Also reported for reference in the right panel is the result for
SCDM.}
\label{fi:j3_lcdm}
\end{figure}

The resulting $J_3(R)$ for Abell/ACO clusters (filled circles) are
compared in Figures \ref{fi:j3_chdm} and \ref{fi:j3_lcdm} to
$\Omega_0=1$ and $\Lambda$CDM models, respectively (open symbols). The
reported simulation errorbars correspond to the $1\sigma$ cosmic
scatter over the ensemble of 48 model observers.
From this plot one may judge the reliability of a
model according to the distance between the Abell/ACO data and the
simulation errorbars.

In order to provide a more quantitative measure of the reliability of a
model, we estimate the probability that the observed Abell/ACO value,
$J_3^{obs}$, is produced by chance, due to cosmic variance fluctuations.
To this purpose, let $J_3^i$ be the measure of the $i$--th observer ($i=1,
\dots ,48$) and $\bar J_3={1\over 48}\sum_i J_3^i$ be the average value. Then,
let ${\cal N}$ be the number of observers for which
$J_3^i\ge J_3^{obs}$ ($J_3^i\le J_3^{obs}$) in a DM model for which
$\bar J_3\le J_3^{obs}$ ($\bar J_3\ge J_3^{obs}$). Therefore, the smaller
the number ${\cal N}$ of such observers,
the smaller the probability that a measure as
discrepant as $J_3^{obs}$ is produced by chance, the larger the confidence
level to which the model can be ruled out. The values of ${\cal N}$ at $R=
19.4\hm$ and $35.3\hm$ are reported in Table \ref{t:clus}, along with the
corresponding $J_3(R)$ values (note that these two fiducial scales nearly
correspond to the correlation length and the mean separation for
Abell/ACO clusters). We also show the values of $J_3$ for the Abell/ACO
sample, along with the corresponding uncertainties which are
the 1$\sigma$ scatter over an ensemble of 200 bootstrap resamplings
(e.g., Ling, Frenk \& Barrow 1986).

\begin{table*}[table]
\centering
{\small
\caption[]{Clustering tests.
Columns 2 and 4: value of $J_3$ (in units of $10^3$) at $R=19.4\hm$ and
$35.3\hm$ respectively.
Columns 3 and 5: number ${\cal N}$ of the 48 observers which 
measure a $J_3$ value
at least as different as the observational value, $J_3^{obs}$, from the
simulation observer--averaged value, $\bar J_3$ (see text).
Column 6: the shape parameter $\Gamma$ defined by eq. (\ref{eq:sh}).
Column 7: the effective spectral index $n_{\rm eff}$ at the $R=20\hm$ scale.
Column 8: probability ${\cal P}_{\chi^2}$ for model PDFs, estimated at $R_{sm}=
20\hm$, to have the Abell/ACO PDF as parent distribution.
Column 9: reduced skewness $S_3$ of the
smoothed cluster distribution, with $R_{sm}=20\hm$.}
\tabcolsep 7pt
\begin{tabular}{lcccccccc} \\ \hline \hline
Model & $J_3(19.4)$ & ${\cal N}_{19.4}$ & $J_3(35.3)$ & ${\cal N}_{35.3}$ &
$\Gamma$ & $n_{\rm eff}$ & ${\cal P}_{>\chi^2}$ & $S_3$ \\ \hline \\
Abell/ACO & $4.2\pm 0.9$ & & $10.0\pm 1.9$ & & & & &
$1.8\pm 0.3$ \\ \\
\multicolumn{8}{l}{CDM models} \\
SCDM & $3.1\pm 0.2$ & 1 & $5.0\pm 0.5$ & 0 & 0.47&$-0.61$&$<10^{-3}$
& $1.8\pm 0.3$ \\
LOWH & $3.7\pm 0.2$ & 6 & $6.5\pm 0.6$ & 0 & 0.31&$-0.83$&$<10^{-3}$
& $1.8\pm 0.3$ \\
LOWH$_{0.9}$ & $3.7\pm 0.2$ & 7 & $6.4\pm 0.5$ &0&0.27&$-0.91$&0.002
& $1.8\pm 0.3$ \\ \\
\multicolumn{8}{l}{CHDM models} \\
C$\nu$DM$_{0.1}$ & $3.9\pm 0.2$ & 12 & $6.7\pm 0.5$&0&0.30&$-0.83$&$<10^{-3}$
& $1.7\pm 0.2$ \\
C$\nu$DM$_{0.2}$ & $4.5\pm 0.3$ & 13 & $8.0\pm 0.6$ & 6&0.21&$-0.99$&0.255
& $1.8\pm 0.2$ \\
C$\nu^2$DM$_{0.2}$ & $4.6\pm 0.3$ & 10 &$8.6\pm 0.6$&9&0.19&$-1.07$&0.776
& $1.8\pm 0.3$ \\
C$\nu$DM$_{0.3}$ & $5.0\pm 0.3$ & 4 & $8.9\pm 0.6$ & 9&0.16&$-1.09$&0.667
& $1.8\pm 0.3$ \\
C$\nu$DM$_{0.5}$ & $5.9\pm 0.3$ & 0 & $10.1\pm 0.6$&25&0.11&$-1.22$&0.955
& $1.8\pm 0.3$ \\ \\
\multicolumn{8}{l}{$\Lambda$CDM models} \\
$\Lambda$CDM$_{0.3}$ & $5.1\pm 0.3$&4&$9.6\pm 0.6$&15&0.18&$-1.15$&0.997
& $1.8\pm 0.3$ \\
$\Lambda$CDM$_{0.4}$ & $4.7\pm 0.3$&8&$8.7\pm 0.6$&9&0.20&$-1.09$&0.991
& $1.8\pm 0.3$ \\
$\Lambda$CDM$_{0.5}$ & $4.0\pm 0.2$&18&$7.1\pm 0.6$&2&0.26&$-0.94$&0.212
& $1.8\pm 0.3$ \\
\hline
\end{tabular}
\label{t:clus}
}
\end{table*}

All such results consistently indicate that CDM models produce in general
too small $J_3(R)$ values. Lowering the Hubble parameter to $h=0. 4$ (LOWH)
and slightly tilting the primordial spectral index to $n=0.9$
(LOWH$_{0.9}$) gives some improvements with respect to SCDM only on scales
$R\mincir 20\hm$. On the other hand, for
$R\magcir 30\hm$ the correlations remain too
weak and no observer in these three models measures a $J_3$ value as
large as the Abell/ACO one. As for the Cold+Hot DM models, it turns out that
$0.2\mincir \Omega_\nu\mincir 0.3$ is required to fit the data, quite
independently of whether the hot component is shared by one or two massive
neutrinos. Lowering further $\Omega_\nu$ reproduces the
SCDM results on large scales. Vice versa, larger values of $\Omega_\nu$
generate an excess clustering on scales $R\mincir 30\hm$,
being consistent with real data only on the largest scales.
As for the $\Lambda$CDM case,
we find that all the three considered models are quite consistent with the
observational constraints, with only the $\Lambda$CDM$_{0.5}$ being
disfavoured by its small $J_3$ value on large scales.

\subsection{The PDF statistics}
The probability density function (PDF), $P(\delta)$, for the density
fluctuation field is defined as the probability for that field to assume a
given value $\delta$. Although it provides the lowest--order (i.e.,
1--point) statistical description, it conveys information about moments of
the density field of any order, via
\be
\lb \delta^q\rb \,=\, \int P(\delta)\,\delta^qd\delta
\label{eq:mom}
\ee
with $\lb \delta\rb =0$ by definition. Eq.(\ref{eq:mom}) defines the
variance and the skewness of the $\delta$--field for $q=2$ and $q=3$,
respectively, while $S_3=\lb \delta^3\rb/\lb \delta^2\rb^2$ is the reduced
skewness.

Several models have been proposed for $P(\delta)$ on the basis of
different approaches to the description of non--linear gravitational
clustering (see, e.g., Sahni \& Coles 1995 and references therein).
Several authors attempted recently to apply the PDF statistics to test such
clustering models, using N--body simulations (e.g., Bernardeau \&
Kofman 1995), as well as the actual galaxy (e.g., Kofman et al. 1994) and the
cluster (Plionis \& Valdarnini 1995) distributions.
Here we are only interested in comparing the PDF statistics for the
Abell/ACO sample to that of the simulations to put constraints on DM
models.

\begin{figure}
\mbox{\epsfxsize=14 cm\epsffile{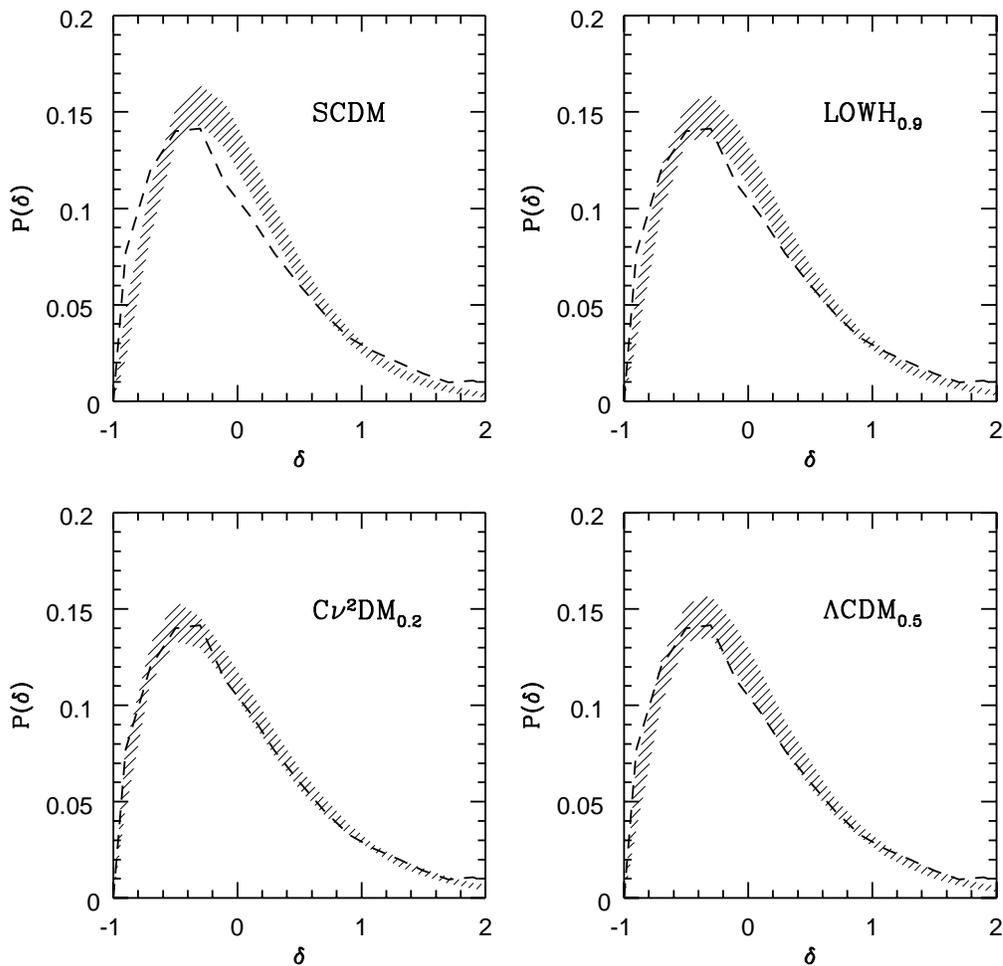}}
\caption
{The probability density function (PDF) for the cluster density
field smoothed with a Gaussian window of $20\hm$ radius for SCDM,
LOWH$_{0.9}$,
C$\nu^2$DM$_{0.2}$ and $\Lambda$CDM$_{0.5}$. The dashed line is the
result for the Abell/ACO sample, while the dashed band is the $1\sigma$
``cosmic" scatter for the simulation PDFs.}
\label{fi:ameri_pdf}
\end{figure}

In order to obtain a continuous field from the discrete cluster
distribution we resort to the same smoothing procedure that has been
applied by Plionis \& Valdarnini (1995; see also Plionis et al. 1995 and
BPCM). It can be summarized as follows.
\begin{description}
\item[(a)] We smooth the cluster distribution by using the Gaussian kernel
\be
{\cal W}(r)\,=\,(2\pi R_{sm}^2)^{-3/2}\exp\left(-{r^2\over 2R_{sm}^2}
\right)\,.
\label{eq:wind}
\ee
\item[(b)] We reassign the density on a grid with $20\hm$ spacing according
to the relation
\be
\rho(\bx_g)\,=\,{\sum_i w_i\,{\cal W}(|\bx_g-\bx_i|)\over
\int {\cal W}(r)\,d^3r}\,.
\label{eq:degr}
\ee
In the above equation,
$\bx_g$ and $\bx_i$ represent the vector positions for the grid point
and for the $i$--th cluster, respectively, while $w_i$ is the weight
assigned to the $i$--th cluster. Since we sum up over clusters
such that $|\bx_g-\bx_i |\le 3R_{sm}$, the integral in the denominator
takes a value slightly smaller than unity ($\simeq 0.97$).
\item[(c)] We consider only those grid points whose position corresponds
to a value of the completeness factor, $f(\bx_g) \ge 0.8$, where
$f(\bx_g) \approx [P(z)\,P(b)]^{-1}$; for the proper
definition of $f$ see Plionis \& Valdarnini (1995).
\end{description}

It is clear that, since the parent cluster distribution has a discrete
nature, the statistics of the smoothed field will be affected by
shot--noise effects. However, we prefer not to correct for such effects.
Firstly, the usual shot--noise corrections (e.g., Peebles 1980)
are based on the assumption that the point distribution
represents a Poisson sampling of an underlying continuous field whose
statistics one is wishing to recover. This is clearly not the case for the
distribution of clusters, which are instead expected to trace the high
density peaks of the DM density field (see e.g. Borgani et al. 1994).
Secondly, our simulation cluster samples
are constructed to have the same number density as the Abell/ACO sample.
Therefore, shot--noise should affect similarly the simulation
and real data analysis, and thus we will be comparing like with
like.

Furthermore the smoothing process itself suppresses significantly
the shot--noise effects (Gazta\~naga \& Yokoyama 1993) and
thus we will consider only $R_{sm}\ge 20\hm$ for the smoothing
radius. BPCM showed that the PDF statistics at $R_{sm}=20
\hm$ are effective in discriminating between different DM models, while the
weak cluster clustering makes such differences hardly detectable at
$R_{sm}=30\hm$. For this reason, we will present in the following only
results based on the smaller $R_{sm}$ value. As an example, we plot in
Figure \ref{fi:ameri_pdf} the PDFs for some models and compare them with the
Abell/ACO results (dashed curve). For model PDFs, the shaded band
represents the 1$\sigma$ scatter estimated over the 48 observers. Between
the four plotted models, it is apparent that the SCDM is the most discrepant;
the weakness of the clustering produced by this model turns into a PDF
shape which is significantly broader that the Abell/ACO one.

In order to provide a quantitative comparison, we also estimate the
quantity
\be
\chi^2\,=\,\sum_i\left[{P^{sim}(\delta_i)-P^{obs}(\delta_i)\over
\sigma_i}\right]^2\,,
\label{eq:chi2}
\ee
where the index $i$ runs over the bins in  $\delta$ and
$\sigma_i^2$ is the cosmic variance in the determination of $P^{sim}(\delta
)$. In Table \ref{t:clus} we report the probability ${\cal P}_{>\chi^2}$
that the simulation PDF could have been drawn by chance from a parent
distribution given by the observational PDF.
The results confirm what we found from the $J_3$
analysis: all the CDM models with $\Omega_0=1$ and the
C$\nu$DM$_{0.1}$ one are rejected at a
quite high confidence level. The only exception is the
C$\nu$DM$_{0.5}$ model, which is now perfectly acceptable. However, these
results are obtained for a Gaussian window of radius $R_{sm}=20\hm$, which
roughly corresponds to $R\simeq 35\hm$ for the radius of the top--hat
sphere on which the $J_3$ test is based. Therefore, this result is
consistent with the agreement between Abell/ACO and C$\nu$DM$_{0.5}$
analyses at that scale.

As in BPCM we have found that the value of the reduced skewness $S_3$ (see
Column 4 of Table \ref{t:clus}), is remarkably independent of the model and
virtually identical to that characterizing the Abell/ACO clusters (see also
Gazta\~naga, Croft \& Dalton 1995; Jing, B\"orner \& Valdarnini 1995).
This indicates that $S_3$ is not
connected to the fluctuation spectrum but either to the nature (i.e.,
Gaussian vs. non--Gaussian) of the initial fluctuations or to the high--peak
biasing prescription for identifying clusters.
In any case, the remarkable agreement between data and all the model
predictions indicates that the Abell/ACO clusters are consistent with being
high density peaks of a mildly non--linear density field, evolved according
to the gravitational instability picture from random--phase initial
conditions (cf. also Kolatt, Dekel \& Primack 1996).

\subsection{The percolation analysis}

Different but complementary information can be obtained by using a
percolation technique, essentially based on the friends--of--friends algorithm.
In fact this method allows one to individuate the largest structures present in
the cluster samples, roughly corresponding to supercluster sizes.

\begin{figure}
\mbox{\epsfxsize=12 cm\epsffile{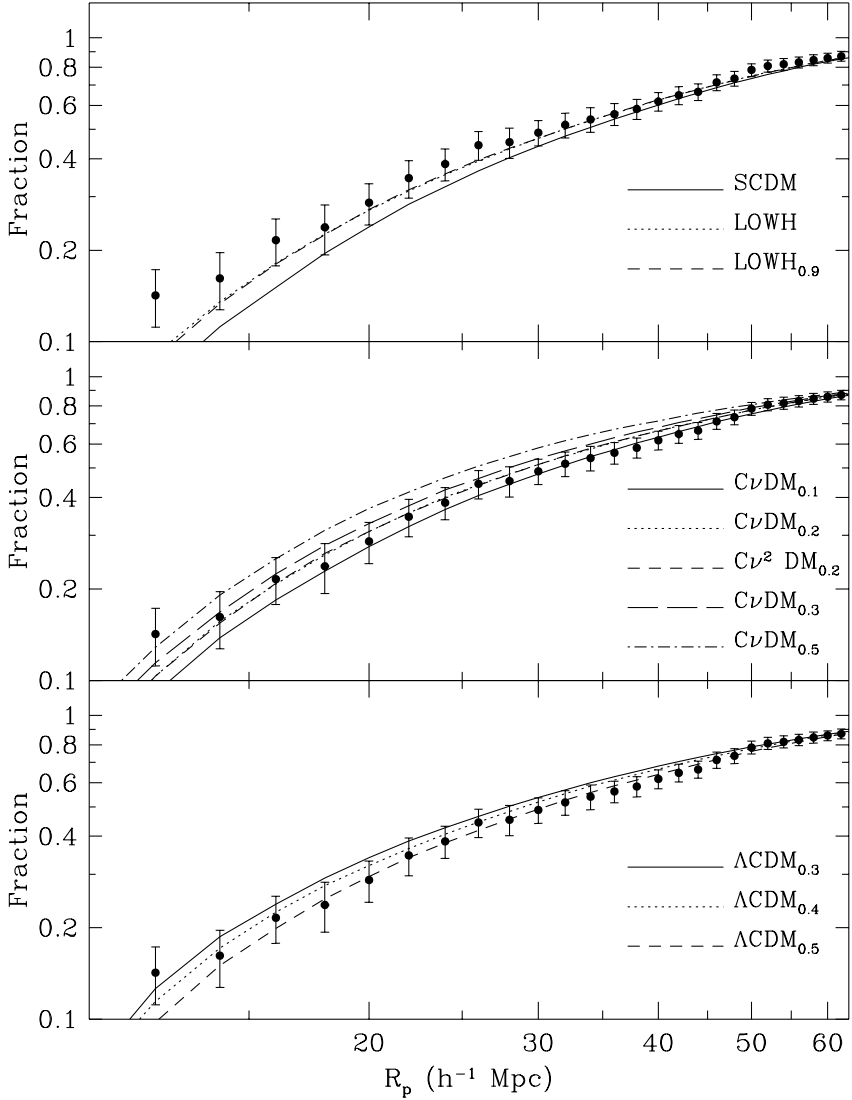}}
\caption{The fraction of clusters in superclusters as a function of the
percolation radius $R_p$ (in $\hm$). The filled circles refer to the
Abell/ACO clusters, while the corresponding values for the different models
are represented by the different lines. The errorbars, centred on the
real data, are the $2\sigma$ scatter estimated over the ensemble of
observers for the SCDM model. The panels refer to the CDM models (top),
the CHDM models (centre) and the $\Lambda$CDM models (down).}
\label{fi:fract}
\end{figure}

Percolation, as a discriminatory tool between different cosmological models,
has been criticized due to the dependence of the results on the mean density of
the objects and/or on the volume sampled (e.g. Dekel \& West
1985). For these reasons we checked the stability of the following results
against possible small changes in the parameters of the simulated catalogues
(the mean separation $d_{cl}$, the ratio between the number density of Abell
and ACO cluster samples, the parameters of the selection function, etc.).
Although, by varying these parameters the final number of clusters inside the
simulated catalogues can vary up to 30\%,  we found that all the main
results are robust, showing variations smaller than the observer--to--observer
variance.

As a first statistical test, we compute the fraction of clusters in
superclusters as a function of the percolation radius $R_p$ (Postman et al.
1989; Plionis, Valdarnini \& Jing 1992, hereafter PVJ), where a
supercluster is defined as a group with a number
of members $N_m \ge 2$. The results for
the different cosmological models are shown in Figure \ref{fi:fract}, where
the behaviour of this statistics for the real sample is also plotted (filled
circles). The errorbars, centred on the real data, are the $2\sigma$ 
errors in the
observer' ensemble for the SCDM model: very similar error sizes are
obtained for the other models. The results are in good agreement with those
obtained by our previous analyses. The SCDM model has a small fraction of
clusters
inside the superclusters for percolation radii $R_p \mincir 30\hm$.
This agrees with the results of the correlation analysis, which indicates a
lack of close cluster pairs in the SCDM model. On the other hand, the
Cold+Hot DM models with $\Omega_\nu=0.5$ forms superclusters too easily,
with a consequent too large fraction of clusters in
superclusters. The other models are all well inside the $2\sigma$ errorbars,
with some exceptions at very small percolation radii ($R_p \le 12 \hm$). This
discrepancy is essentially due to the fact that our simulation method is not
suitable to resolve the very close pairs of clusters, reducing
systematically the percolation at these small radii.

As a further test, we compute the multiplicity function MF
(i.e. the number of superclusters with a given cluster membership $N_m$)
at two different percolation radii, $R_p=20$ and
$35 \hm$: these two values have been chosen to approximate the cluster
correlation length and the mean intercluster separation, respectively. In
order to decrease the shot--noise effects and thus to perform a reliable
statistical analysis, we prefer to bin our results in logarithmic intervals;
we tested that the final results are almost independent of the choice of the
binning parameters. We compare the MF of the real
sample to that of each simulated catalogue by using a $\chi^2$--test. In Table
\ref{t:perc} we report, for both percolation radii, the fraction
${\cal F}_{MF}$ of the 480
simulated catalogues (corresponding to the 10 selection function realizations
for each of the 48 observers) of each model
with a $\chi^2$ probability of reproducing the real data which is
larger than 0.15. Although some differences exist between the models, the
fraction of such ``good'' observers is very large for all the models ($>0.8$)
which indicates that the MF is not a discriminatory test.
As a consequence, all the models agree with the observational data.

The shapes of the superclusters have been suggested as a useful tool to
discriminate among different models (see, e.g., Matarrese et al. 1991; PVJ).
This statistics is based on the computation for each supercluster
of the inertia tensor $I_{kl}=
\sum_{i=1}^{N} (X_k X_l)_{_i} W_{_i}$, where the indices $k$ and $l$ range from
1 to 3, and $X_{1_i}$, $X_{2_i}$, $X_{3_i}$ are the cartesian coordinates of
the $i$--th cluster computed with respect to the centre of mass of
the supercluster.
In order to reduce the importance of the outlying clusters, we weight
each cluster by $W_{_i}=(X_{1_i}^2+X_{2_i}^2+X_{3_i}^2)^{-1}$, which is
the inverse square cluster distance from the supercluster centre of
mass (West 1989; PVJ). We then compute the following three quantities,
$\lambda_1=1/a_3^2$,
$\lambda_2=1/a_2^2$ and  $\lambda_3=1/a_1^2$, where $a_1^2 \ge a_2^2 \ge a_3^2$
are the eigenvalues of the principal axes of the inertia tensor. Following
Bardeen et al. (1986), we define for each supercluster two shape parameters:
\be
{e} = {{\lambda_1 - \lambda_3} \over {2 \sum_{i=1}^{3} \lambda_i}} \ge 0
\ee
and
\be
{p} = {{\lambda_1 - 2\lambda_2 + \lambda_3} \over {2 \sum_{i=1}^{3}
\lambda_i}}.
\ee
The quantity $e$ is a measure of the ellipticity in the $\lambda_1-\lambda_3$
plane while $p$ measures the prolateness (if $-e \ge p \ge 0$) or oblateness
(if $0 \ge p \ge e$) of the supercluster. The limiting cases $p=-e$ and $p=e$
represent the prolate and oblate spheroid, respectively.

\begin{figure}
\mbox{\includegraphics{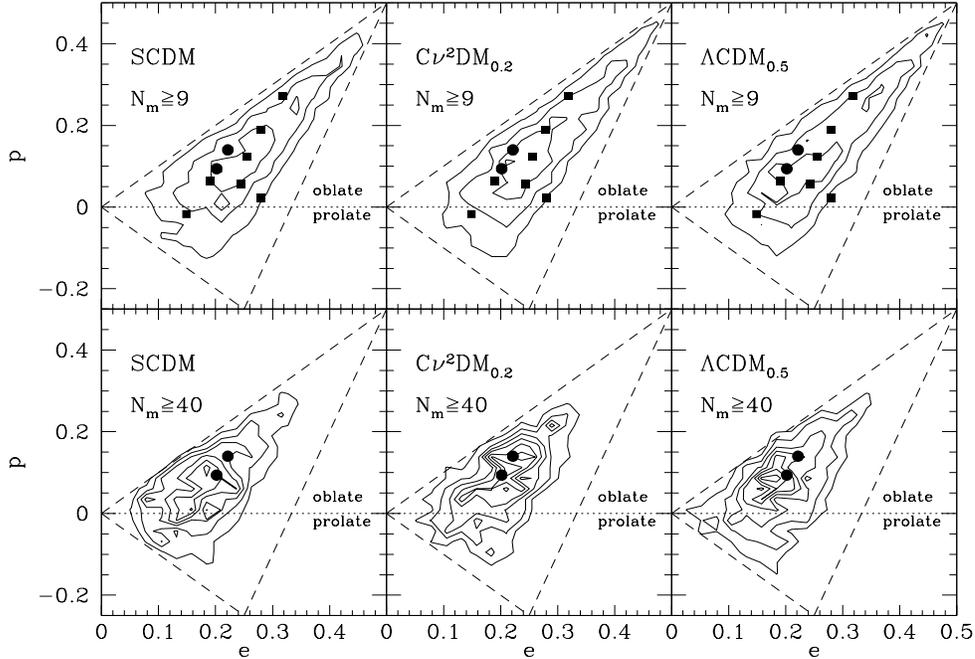}}
\vspace{10.truecm}
\caption{Comparison of the shape parameters $e$ and $p$ for the observed and
the simulated superclusters
found with a percolation radius $R_p=35\hm$.
Only the SCDM (left panels), C$\nu^2$DM$_{0.2}$ (central panels) and
$\Lambda$CDM$_{0.5}$ (right panels) models are reported here. Upper and lower
panels are for
superclusters with $N_m \ge 9$ and $N_m\ge 40$, respectively.
The dashed lines represent the triangle where the values of $e$ and $p$ are
constrained to lie.
The solid lines refer to the isoprobability contours: the spacing is 0.004
starting from the external contour referring to a probability of 0.004.
The real data are represented by filled circles, if the number of
members is $N_m\ge 40$, or filled squares, if $9\ge N_m < 40$.}
\label{fi:oblate}
\end{figure}

In order to have reliable statistical characterization of supercluster
shapes, we need at the same time a
large number of such structures, each containing a rather large number
of members. For this reason, we consider in this analysis all the
simulated superclusters with $N_m \ge 9$. In Figure \ref{fi:oblate}
we show the distributions of values of $e$ and $p$ obtained
for SCDM, C$\nu^2$DM and $\Lambda$CDM$_{0.5}$ models,
when a percolation radius of $R_p=35\hm$ is adopted (similar results have been
obtained for the other models and are not reported). The
solid lines represent the isoprobability contours while the dashed lines show
the triangle where the values of $e$ and $p$ are constrained to lie. The
results for the nine Abell/ACO superclusters are
displayed as filled symbols (in the upper panels, circles are for
the two superclusters with $N_m\ge
40$, squares for $N_m < 40$). The behaviour for the different models is very
similar: the shapes are systematically triaxial,
with a tendency to oblateness. Moreover we find, in agreement with previous
analyses (Matarrese et al. 1991; PVJ) and with theoretical predictions
for a Gaussian random field (e.g., Bardeen et al. 1986), that
the richer superclusters have the tendency to be more spherical than the poorer
ones. In general we
find that the parameter values obtained for the real superclusters are rather
typically reproduced by all models.
In Table \ref{t:perc} we report the fraction ${\cal F}_{obl}$ of
superclusters having an oblate shape, i.e. with
$p\ge 0$, computed for a percolation radius of both 20 and $35\hm$, for both
real and simulated samples. We note that, since a total number of only 6
superclusters characterizes the statistics of the Abell/ACO sample, the
corresponding Poissonian uncertainties are so large that
all the models are in agreement with the results for the real sample.

Furthermore, we also report in Table \ref{t:perc} the number of
clusters inside the
richest supercluster, $N_m^{max}$, computed for two values of the
percolation radius, $R_p=20$ and $35\hm$, along with the
$1\sigma$ scatter within the observer' ensemble. At the smaller $R_p$,
the only models which do not agree with the Abell/ACO result are the SCDM,
both the LOWH models and, more marginally, C$\nu$DM$_{0.1}$, which
produce too poor superclusters. On the other hand, the much larger cosmic
variance at $R_p=35\hm$ does not allow to detect any significant difference
between the models.

Therefore, combining the results about the multiplicity function with those
from $N_m^{max}$, we conclude that Cold+Hot DM models with
$\Omega_\nu=0.2$--0.3 and the considered $\Lambda$CDM models reproduce rather
well the percolation properties of the Abell/ACO cluster sample, thus in
agreement with the constraints obtained from the previous analyses.

\begin{table*}[table]
\centering
\caption[]{Percolation  tests.
The fraction ${\cal F}_{MF}$ of simulated catalogues having a
$\chi^2$  probability of reproducing the observed multiplicity
function larger than 0.15; the fraction ${\cal F}_{obl}$ of superclusters
having an oblate shape; the number of clusters $N_m^{max}$ inside the
richest supercluster.}
\tabcolsep 7pt
\begin{tabular}{lcccccc} \\ \hline \hline
Model     & ${\cal F}_{MF}$ & ${\cal F}_{obl}$ & $N_{m}^{max}$ &
            ${\cal F}_{MF}$ & ${\cal F}_{obl}$ & $N_{m}^{max}$ \\
& \multicolumn{3}{c}{perc. radius $R_p=20\hm$} &
  \multicolumn{3}{c}{perc. radius $R_p=35\hm$} \\ \hline \\
Abell/ACO & & 0.667 & 17 & & 0.889 & 85 \\ \\
\multicolumn{6}{l}{CDM models} \\
SCDM & 0.817 & 0.854 & $11.3\pm  2.3$ &
       0.735 & 0.830 & $93.4\pm 31.7$ \\
LOWH & 0.873 & 0.884 & $12.9\pm  3.1$ &
       0.792 & 0.820 & $92.1\pm 36.2$ \\
LOWH$_{0.9}$ & 0.896 & 0.879 & $12.9\pm  2.8$ &
               0.786 & 0.839 & $96.0\pm 36.3$ \\ \\
\multicolumn{6}{l}{CHDM models} \\
C$\nu$DM$_{0.1}$ & 0.915 & 0.867 & $13.2\pm  3.6$ &
                   0.746 & 0.820 & $90.7\pm 33.3$ \\
C$\nu$DM$_{0.2}$ & 0.933 & 0.879 & $14.7\pm  4.8$ &
                   0.819 & 0.832 & $89.8\pm 28.5$ \\
C$\nu^2$DM$_{0.2}$ & 0.980 & 0.881 & $15.4\pm  5.1$ &
                     0.800 & 0.837 & $92.3\pm 30.2$ \\
C$\nu$DM$_{0.3}$ & 0.954 & 0.888 & $15.3\pm  4.0$ &
                   0.813 & 0.835 & $89.6\pm 29.5$ \\
C$\nu$DM$_{0.5}$ & 0.960 & 0.876 & $18.1\pm  5.1$ &
                   0.798 & 0.854 & $86.8\pm 27.4$ \\ \\
\multicolumn{6}{l}{$\Lambda$CDM models} \\
$\Lambda$CDM$_{0.3}$ & 0.931 & 0.882 & $16.9\pm  4.7$ &
                       0.860 & 0.844 & $104.4\pm 41.3$\\
$\Lambda$CDM$_{0.4}$ & 0.921 & 0.868 & $15.4\pm  4.4$ &
                       0.829 & 0.851 & $101.0\pm 37.5$\\
$\Lambda$CDM$_{0.5}$ & 0.865 & 0.882 & $14.5\pm  3.8$ &
                       0.827 & 0.830 & $97.4\pm 31.4$ \\
\hline
\end{tabular}
\label{t:perc}
\end{table*}

\section{Discussion}
\subsection{Spectrum shape}
The results that we presented in the previous section converge to indicate
that the cluster distribution is indeed a powerful tool to constrain
different power spectrum models. The results reported in Table
\ref{t:clus} show that the statistics of the cluster
distribution is quite sensitive to the power spectrum shape, while
independent analyses (e.g., Croft \& Efstathiou 1994; BPCM) show that the
amplitude of the cluster correlations in a given model is almost independent of
the normalization or evolutionary stage.

In order to understand in more detail the constraint that the cluster
distribution provides on the spectrum profile of cosmological 
models, we decided
to correlate the cluster $J_3$ values with suitable parameters
defining the profile of $P(k)$. In particular, we considered the shape
parameter defined by
\be
\Gamma\,=\,0.5\,\left({0.95\sigma_8\over 3.5\sigma_{25}}\right)^{1/0.3}\,
=\,0.3826 (EP)^{-1/0.3}\,,
\label{eq:sh}
\ee
($\sigma_8$ and $\sigma_{25}$ are the rms fluctuation amplitudes within
top--hat spheres of $8\hm$ and $25\hm$ radii, respectively) and the
effective spectral index
\be
n_{\rm eff}\,=\,-3-{d\log \sigma^2(R)\over d\log R}\,.
\label{eq:neff}
\ee
With the definition (\ref{eq:sh}), for SCDM and the $\Lambda$CDM models
the shape parameter is approximately
equivalent to the quantity $\Gamma=\Omega_0 h\exp(
-\Omega_{\rm bar}-\Omega_{\rm bar}/\Omega_0)$, which specifies the CDM transfer
function (Efstathiou, Bond \& White 1992), suitably corrected to account
for a non--negligible baryon contribution (Peacock \& Dodds 1994; Sugiyama
1995).  In eq. (\ref{eq:sh}), the ``Excess Power'' $EP \equiv 3.4
\sigma_{25}/\sigma_8$, defined so that SCDM has $EP \approx 1$, was
introduced by Wright et al. (1992), who stated
that $EP \approx 1.3$ is needed to fit the APM galaxy angular correlation
function; with the above definition, this corresponds to $\Gamma \approx 0.16$.

\begin{figure}
\vspace{-5.truecm}
\mbox{\epsfxsize=14 cm\epsffile{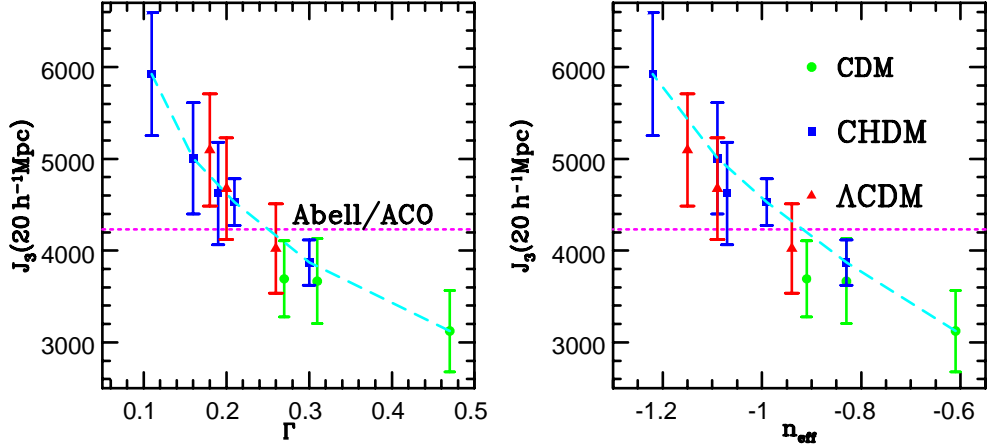}}
\caption{The value of the $J_3$ integral for all the models evaluated at
$R=20\hm$ is plotted against the value of the corresponding shape
parameter $\Gamma$ (left panel) and effective spectral index, $n_{\rm eff}$
evaluated at the same scale. The horizontal dotted line is the $J_3$
value for the Abell/ACO sample.}
\label{fi:shape}
\end{figure}

We plot in Figure \ref{fi:shape} the $J_3$ values at $R=20\hm$
for each model versus
the corresponding $\Gamma$ (left panel) and $n_{\rm eff}$ (right panel) values.
We decided not to
include plots of results for larger scales, for which we have already 
verified that the results
are less discriminatory. The values of $\Gamma$ and $n_{\rm eff}$ for all the
models have been already reported in Table \ref{t:clus}. Apart from the
individual model details, the correlation between the spectrum shape and
the $J_3$ values is really striking: the steeper the spectrum, the
stronger
the cluster correlations, with the extreme two models being C$\nu$DM$_{0
.5}$ and SCDM. 
The trends are especially clear if one compares models of the same
class, represented in Figure \ref{fi:shape} by points of the same
shape.  The dashed line connects the four C$\nu$DM models and SCDM,
obtained by varying $\Omega_\nu$ from 0.5 to 0.  The C$\nu^2$DM model
lies only slightly below this dashed line.  The fact that $J_3$ shows
the same trend as a function of $n_{eff}$ or of our generalized
$\Gamma$ shape parameter shows that these characterize well the
behaviour of the cluster correlations.  

A comparison with the $J_3$ result for the Abell/ACO sample
(horizontal dotted lines) leads to $0.18\mincir \Gamma \mincir 0.25$ and 
$-1.1\mincir n_{\rm eff}\mincir -0.9$ at $R\simeq 20\hm$ being
required by the cluster distribution. It is interesting to note 
that this value of
$\Gamma$ agrees with that, $\Gamma =0.23\pm 0.04$, obtained by Peacock \&
Dodds (1994) from the power spectrum analysis of APM galaxies (cf. also
Viana \& Liddle 1995). This suggests that optical galaxies and Abell/ACO
clusters actually trace the same power spectrum, at least on such scales,
the only difference arising from their different linear biasing factors. 
Recently Eke et al. (1996b) considered SCDM cluster N--body 
simulations in order to check the stability of 
the cluster clustering when different prescription for their identification 
are applied. Although they found rather marginal differences, it is clear 
that realistic cluster identification methods in simulations should be 
considered when new observational cluster samples based on improved selection 
criteria will be available.

\subsection{Spectrum amplitude}
Although the statistics of the cluster distribution represents a
useful tool for providing hints about the shape of the power spectrum,
nevertheless one may well imagine that different combinations of the
fundamental parameters which enter in defining a model (i.e., DM composition,
values of $\Omega_0$, $\Omega_\Lambda$ and $h$) could produce the
same profile of the transfer function. For this reason, further
observational tests are required to constrain also the fluctuation
amplitude, which is expected to vary from model to model, due to the
different way in which the defining parameters enter in specifying the
large--scale normalization to CMB temperature fluctuations. The most
constraining of such tests is probably the abundance of
galaxy clusters (see, e.g., White, Efstathiou \& Frenk 1993; BPCM;
Viana \& Liddle 1995; Eke, Cole \& Frenk 1996a; 
Borgani et al. 1996a,b, and references therein).

Since our TZA simulations are inadequate to describe mass accretion
into clusters, we resort here to the analytical approach provided by Press
\& Schechter (1974, PS). According to this recipe, the number density of
collapsed structures of mass larger than $M$ is
\be
N(>M)\,=\,\int_M^\infty n(M')\,dM'\,.
\label{eq:cum}
\ee
Here, $n(M)\,dM$ is abundance of structures with mass
between $M$ and $M+dM$, which reads
\be
n(M)\,dM\,=\,\sqrt{2\over \pi}\,{\bar \rho \delta_c\over \sigma_M}\,
\bigg|{d\log \sigma_M\over d\log M}\bigg|\,\exp\left(-{\delta_c^2\over
2\sigma_M^2}\right)\,{dM\over M^2}\,,
\label{eq:prsc}
\ee
where $\bar \rho$ is the average cosmic density, $\delta_c$ is the critical
density contrast for a linearly evolved overdensity to turn into a collapsed
structure, and
\be
\sigma_M\,=\,\left[{1\over 2\pi^2}\,\int P(k)\,W^2(kR)\,k^2dk\right]^{1/2}
\label{eq:sigm}
\ee
is the rms mass fluctuation within the window $W(kR)$. As usual, the mass
$M$ is connected to the window radius $R$ according to $M=fR^3$, $f$ being
a form factor, which depends on the $W$ profile. For the Gaussian window
$W(kR)=\exp(-k^2R^2/2)$ that we will consider in
the following, it is $f=(2\pi) ^{3/2}$.

As for the $\delta_c$ value, linear theory for spherical top--hat collapse
predicts $\delta_c=1.69$ (e.g., Peebles 1980). However, deviations from
this value are expected on several grounds. For instance,
asphericity of the collapse would in general predict a lower $\delta_c$
(e.g., Monaco 1995). Such an effect is generally believed to be of limited
relevance, since exceptionally high peaks are expected to closely
follow a spherical collapse (e.g., Bernardeau 1994). It is however not
clear whether this high--peaks approximation would apply to real clusters.
Indeed, for models with $\sigma_8\simeq 0.8$ and $\delta_c\simeq 1.7$,
clusters would arise just from $\simeq 2\sigma$ overdensities of the initial
linear field. From N--body simulations of the C$\nu$DM$_{0.3}$
model, Klypin \& Rhee (1994) and Walter \& Klypin (1996) found that 
eq.(\ref{eq:prsc}) with
$\delta_c\simeq 1.5$ and Gaussian window provides a good fit to the
resulting cluster mass function. Although the stability of this result when
changing the DM model should be still understood in detail, it is well
established that linear--theory predictions for $\delta_c$ do not
depend strongly on the value of $\Omega_0$ (e.g., Lilje 1992; Lacey \& Cole
1993; Colafrancesco \& Vittorio 1994). This has been been confirmed by the
N--body approach followed by White et al. (1993) for CDM--like models with
$\Omega_0=0.2$ and $\Omega_0=1$ (cf. also Mo et al. 1996).

\begin{figure}
\mbox{\epsfxsize=14 cm\epsffile{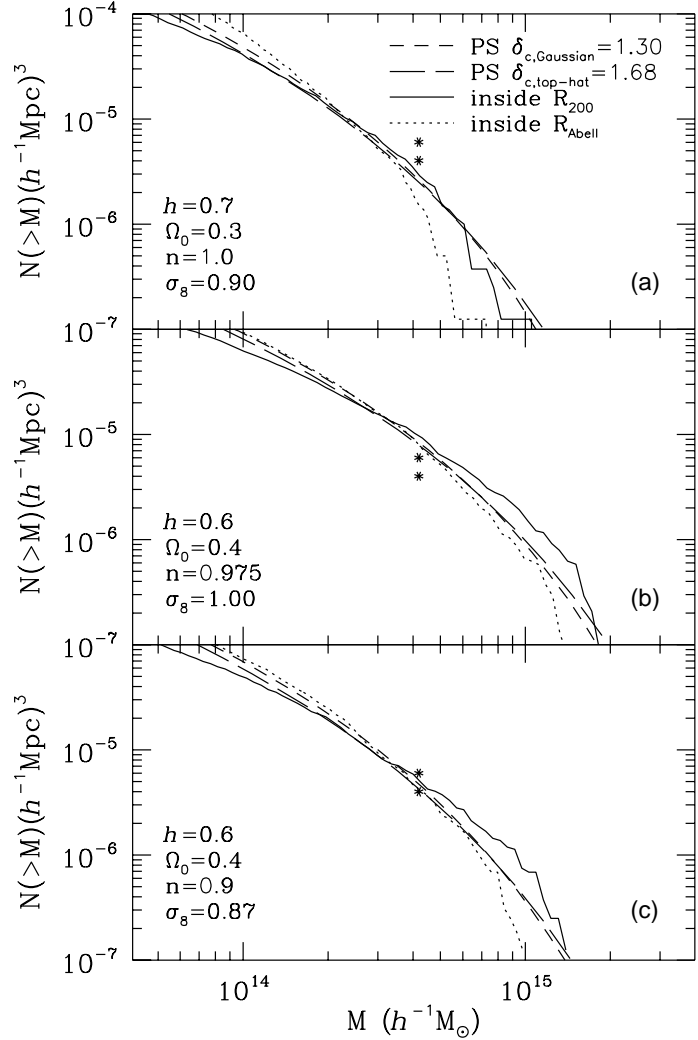}}
\caption{The abundance $N(>M)$ of clusters as a function of mass $M$,
comparing results from Press--Schechter theory with those from N--body
simulations for three $\Lambda$CDM models: (a) $\Omega_0=0.3$,
$h=0.7$, $n=1$, and $\sigma_8=0.9$; (b) $\Omega_0=0.4$, $h=0.6$,
$n=0.975$, and $\sigma_8=1.0$; (c) $\Omega_0=0.4$, $h=0.6$, $n=0.9$,
and $\sigma_8=0.87$.  Dashed lines are for Press-Schechter (PS)
approximations for the models.  The short dashed line represents PS
with a Gaussian filter and $\delta_c=1.3$.  The long dashed line
represents PS with a top-hat filter and $\delta_c=1.68$.  The solid
and dotted lines are for mass functions (MFs) from simulations.  Solid
lines are for MFs that have been constructed from halos defined by the
spherical radius at which the overdensity falls below 200.  Dotted
lines are for MFs that have been constructed from halos defined by the
spherical Abell radius of $1.5 h^{-1}$ Mpc. The upper star symbol on each
panel represents the observational data point of Biviano et al. (1993), 
and the lower one is that of White et al. (1993).  See text for 
discussion.}
\label{fi:simsvsPS}
\end{figure}

We show in Figure \ref{fi:simsvsPS} the results of some new N--body
simulations undertaken just to clarify this point, and also the
dependence on the precise definition of the region regarded as the
cluster, defined either by the Abell radius $1.5 h^{-1}$ Mpc, or by
the radius at which the overdensity falls below 200 (the latter is
larger for rich clusters).  These N--body results are from PM
simulations with a $512^3$ force mesh and $256^3$ particles in a 200
$h^{-1}$ Mpc box.  The mass functions constructed for model (a) are
for just a single realization, while model (b) is an average of three
realizations, and model (c) is the average of two realizations 
\footnote{Note that the normalization of model (a) is lower than that
in Table 1: $\sigma_8=1.06$ there, while $\sigma_8=0.90$ for this
simulation. Such a low normalization has been advocated by some
authors to avoid the excess power on small scales calculated for this
model by Klypin et al. (1996).  But our results in Panel (a) confirm
that the cluster abundance with this lower model normalization is
lower than the observations indicate, according to White et al. (1993)
and Biviano et al. (1993).  Models (b) and (c) are also different from
the $\Lambda$CDM$_{0.4}$ model considered in this paper, since the
spectra are slightly tilted ($n=0.975$ or 0.90 with $T/S=0$).
However, the results are consistent with our PS calculations presented
in Figure 9.  For example, if a PS curve were plotted for model (c) in
Figure 9, it would cross the middle of the dark shaded band at
$\delta_c = 1.3$.  The tilted $\Lambda$CDM$_{0.4}$ model (c) is being
investigated further by Klypin et al. (in preparation) as an example
of a $\Lambda$CDM model that might avoid the problem of too much power
on small scales.}. The
upper and lower asterisks in the figure represent the observational
data points of Biviano et al. (1993) and White et al. (1993).  For
cluster mass in the region of these observational data points, the mass
function fluctuates by about 10\% from the average value from
realization to realization.

From the results shown, it appears that the Press-Schechter
approximation with a Gaussian filter and $\delta_c=1.3$ is a good fit
to the mass function from simulations.
This PS fitting function is very similar
in shape to the one obtained with a top-hat filter and the
canonical $\delta_c=1.68$.  In general, both of these PS curves fit
the mass functions from simulations with the $R_{200}$ definition
reasonably well and the mass functions with the $R_{Abell}$ definition
very well.  The only clear exception is in the case of model (a) where
the $R_{Abell}$ definition is significantly below the PS curves and
the data points.  However, by nature this model has fewer massive
clusters than the others, so the steep decline at the high-mass end of
the curve may be due to statistical fluctuation caused by poor
counting statistics.  Note that model (c), which best fits the
cluster abundance data points, has 
$\sigma_8 \Omega_0^{0.53-0.13(1-\Omega_0)} = 0.56$, thus somewhat 
larger than the value $0.50\pm 0.04$ obtainable for the Eke et al. (1996a)
results for this model.

\begin{figure}
\mbox{\epsfxsize=14 cm\epsffile{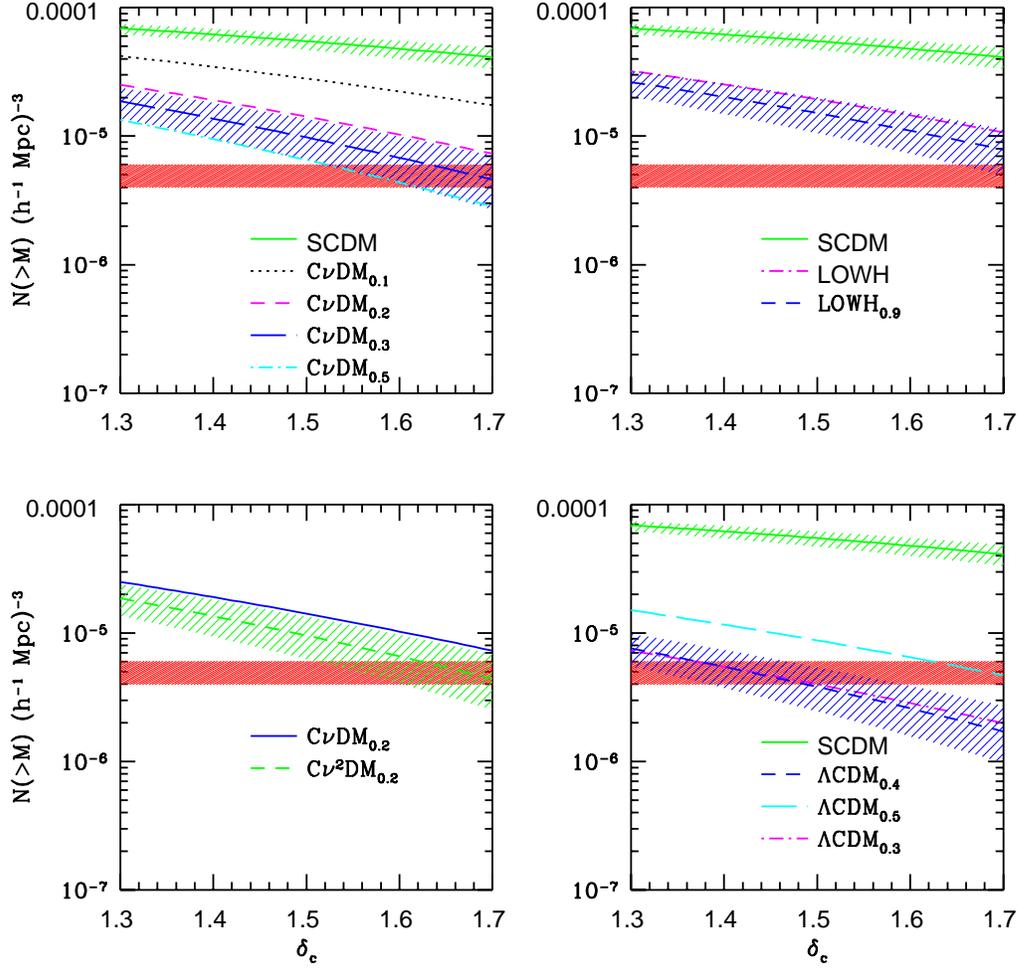}}
\caption{The abundance $N(>M)$ of clusters having mass larger than
$M=4.2\times 10^{14}h^{-1}M_\odot$ arising from the Press--Schechter
theory, plotted as a function of the critical density contrast,
$\delta_c$. Results are obtained by adopting a Gaussian window function.
The dashed band is delimited from below by the observational result of
White et al. (1993) and from above by that of Biviano et al. (1993).}
\label{fi:ameri_ps}
\end{figure}

We can summarize the discussion above by saying that for CHDM models 
there is a good fit to the cluster abundance with $\delta_c=1.5$ in
PS with a Gaussian filter, and for $\Lambda$CDM models $\delta_c=1.3$ 
gives a good fit.
In order to have a rather conservative approach, we prefer to work out
predictions of the cluster abundance for all the models we consider
for a range of $\delta_c$ values (see also BPCM), $1.3\le \delta_c \le
1.7$.  
In Figure \ref{fi:ameri_ps} we plot $N(>M)$ as a function of $\delta_c$
for all the models,
assuming $M=4.2\times 10^{14}h^{-1}M_\odot$. This value for the limiting mass
is that at which White et al. (1993) report cluster abundances from X--ray
and velocity dispersion data. Note that larger values of $M\simeq
10^{15}h^{-1}M_\odot$ would sample the mass function on its high--mass
exponential tail, where any uncertainty in the mass determination would
affect $N(>M)$ more. The shaded band in Figure \ref{fi:ameri_ps} indicates the
observational results; it is delimited from below by the X--ray based abundance
by White et al. (1993), and from above by the determination by Biviano et al.
(1993), based on cluster velocity dispersions. Although realistic observational
uncertainties are probably larger than the difference between such results,
it is not clear whether they can account for the large discrepancies
displayed by some models. For instance, even if cluster masses from X--ray
data were underestimated by a factor of two (cf. Balland \& Blanchard 1995;
see, however, Evrard, Metzler \& Navarro 1996, 
for arguments in favour of a precise mass
determination), the SCDM prediction would decrease to $N(>M)\simeq 1.8\times
10^{-5}$, still quite far from the observational result. The four panels of
Figure \ref{fi:ameri_ps}
correspond to different ways of lowering the SCDM prediction of
$N(>M)$: {\bf (a)} introducing a hot component; {\bf (b)} sharing
$\Omega_\nu$ between more massive neutrinos; {\bf (c)} lowering $h$ and/or
tilting the primordial $P(k)$; {\bf (d)} lowering $\Omega_0$, still keeping
flatness. Note that increasing to some extent the baryon fraction would
also decrease $N(>M)$. In view of such uncertainties in both the observational
outputs and the values of the model parameters, we believe that  it is fair to
draw the following conclusions.
\begin{description}
\item[(a)] Critical density CDM models are allowed only by substantially
lowering the Hubble parameter at least to $h=0.4$ and/or tilting the
primordial spectral index at least to $n=0.9$.
\item[(b)] The Cold+Hot DM model with $n=1$ is
required to have $\Omega_\nu \magcir 0
.3$, with $\Omega_\nu\simeq 0.2$ still allowed if more than
one massive neutrino species share the hot contribution.  A larger
parameter space is allowed if the primordial spectral index is allowed to
tilt slightly below the Zel'dovich value $n=1$ (cf. Primack et al. 1995,
1996; Liddle et al. 1995a).
\item[(c)] $\Lambda$CDM models should have $\Omega_0\mincir 0.4$ in
order to provide a substantial improvement with respect to SCDM for $n=1$,
with a larger parameter space available for $n<1$ (Klypin et al. 1996;
Liddle et al. 1995b).
\end{description}

\section{Conclusions}
In this paper we discussed extensively how the distribution of galaxy clusters
can be used to constrain cosmological models for large--scale structure
formation. Our reference cluster distribution is a
combined redshift sample of Abell/ACO clusters. We compare this observational
distribution to that predicted by a list of eleven DM models. To
this purpose, we run for each model a large set of numerical simulations based
on an optimized version of the truncated Zel'dovich approximation (TZA).
Taking advantage of the reliability and flexibility of this simulation method,
we end up with 48 independent mock samples for each model, each reproducing
the same cluster number density and selection functions as the Abell/ACO one.
We use this ensemble of realizations to reliably estimate cosmic variance
effects.

In order to assess the reliability of the considered models, we apply to both
real and simulated samples several clustering analyses, like the correlation
statistics based on the $J_3$ integral, the probability density function and
supercluster statistics based on percolation method. The conclusions that we
draw from these analyses can be summarized as follows.
\begin{description}
\item[(a)] All the $\Omega_0=1$ CDM models produce too weak cluster
clustering. Lowering the Hubble parameter to $h=0.4$ and/or
tilting the primordial spectral index to $n=0.9$ does not improve the
standard CDM model enough.  This is true regardless of the normalization
$\sigma_8$.
\item[(b)] Cold+Hot DM models with $0.2\mincir \Omega_\nu\mincir 0.3$ fit
the Abell/ACO data rather well, almost independent of whether the hot component
is shared between one or two massive neutrino species.
A model with $\Omega_\nu = 0.1$ produces weak clustering, similar to the
SCDM model. On the contrary, the $\Omega_\nu =0.5$ case fails in that
it overproduces clustering on scales $\mincir 30\hm$ and makes clusters
percolate far easier than is observed.
\item[(c)] All the three considered $\Lambda$CDM models fare rather well, with
the $\Omega_0=0.5$ case being less favoured due to the marginal lack of
clustering at $R\simeq 35\hm$.
\item[(d)] As a general result, we find that the cluster distribution
presents a constraint only for the power--spectrum shape, but not for its
amplitude. Reproducing the observed Abell/ACO cluster clustering requires
$0.18\mincir \Gamma\mincir 0.25$ for the shape parameter (\ref{eq:sh}),
or $-1.1\mincir n_{\rm eff}\mincir -0.9$ for the effective spectral index
(\ref{eq:neff}) at the $20\hm$ scale.
\end{description}

In order to constrain also the power spectrum amplitude, we resort to the
estimation of the cluster abundance, which is essentially determined by
$\sigma_8$, the rms linear fluctuation value within a top--hat sphere of
$8\hm$ radius (e.g., White et al. 1993). Since no reliable mass
determination is allowed by our TZA cluster simulations, we use the Press
\& Schechter (1974) approach. We think of the Press--Schechter formula as
a parametric fitting to the ``true'' mass function, where our ignorance
about the details of the non--linear gravitational clustering is accounted
for by varying the critical density parameter $\delta_c$ in a rather broad
range of values.  Comparing to N--body simulations, good fits to the
cluster mass function are obtained for PS with gaussian filter with
$\delta_c \approx 1.5$ for CHDM models and $\delta_c \approx 1.3$ for
$\Lambda$CDM models.

Combining results from the analysis of the spatial distribution and the
abundance of clusters it turns out that the only viable models, of those
considered here, are either the Cold+Hot DM ones with $0.2\mincir
\Omega_\nu \mincir 0.3$, better if shared between more than one massive
$\nu$ species, or $\Lambda$CDM models with $0.3\mincir \Omega_0 \mincir
0.5$ and Hubble constant tuned in such a way to give about 14 Gyrs for the
age of the Universe.

\section*{Acknowledgments.}
SB and LM have been partially supported by Italian MURST. AK, JRP, and CLS
acknowledge support from NASA and NSF grants, and the use of the CM5
at the National Center for Supercomputer Applications, University of
Illinois, Urbana-Champaign.

\section*{References}

Abell G.O., 1958, ApJ, 3, 211 \\
Abell G.O., Corwin H.G., Olowin R.P., 1989, ApJS, 70, 1\\
Achilli S., Occhionero F., Scaramella R., 1985, ApJ, 299, 577\\
Adams F., Bond J.R., Freese K., Frieman J.A., Olinto A., 1993,
Phys. Rev. D, 47, 426\\
Athanassopoulos C., et al., 1995, Phys. Rev. Lett., 75, 2650\\
Bahcall N.A., 1988, ARA\&A, 26, 631\\
Bahcall N.A., Cen R., 1992, ApJ, 398, L81\\
Bahcall N.A., Soneira R.M., 1983, ApJ, 270, 20\\
Balland C., Blanchard A., 1995, preprint astro-ph/9510130\\
Bardeen J.M., Bond J.R., Kaiser N., Szalay A.S., 1986, ApJ,
304, 15\\
Batuski D.J., Bahcall N.A., Olowin R.P., Burns J.O., 1989,
ApJ, 341, 599\\
Bennett C.L., et al., 1996, ApJ, 464, L1\\
Bernardeau F., 1994, ApJ, 427, 51\\
Bernardeau F., Kofman L., 1995, ApJ, 443, 479\\
Biviano A., Girardi M., Giuricin G., Mardirossian F., Mezzetti
M., 1993, ApJ, 411, L13\\
Blumenthal G.R., Dekel A., Primack J.R., 1988, ApJ, 326, 539\\
Blumenthal G.R., Faber S.M., Primack J.R., Rees M.J., 1984,
Nature, 311, 517\\
Bonometto S.A., Valdarnini R., 1984, Phys. Lett., A103, 369\\
Borgani S., 1990, A\&A, 240, 223\\
Borgani S., Coles P., Moscardini L., 1994, MNRAS, 271, 223\\
Borgani S., Coles P., Moscardini L., Plionis M., 1994, MNRAS,
266, 524\\
Borgani S., Gardini A., Girardi M., Gottl\"ober S., 1996b, in preparation\\
Borgani S., Lucchin F., Matarrese S., Moscardini L., 1996a, MNRAS,
280, 749\\
Borgani S., Plionis M., Coles P., Moscardini L., 1995, MNRAS,
277, 1191 (BPCM)\\
Branch D., Nugent P., Fisher A., 1995, Proceedings of the NATO
Advanced Study Institute on Thermonuclear Supernovae,
preprint astro-ph/9601006\\
Cen C.J., Gnedin N.Y., Kofman L.A., Ostriker J.P., 1992, ApJ, 399,
L11\\
Colafrancesco S., Vittorio N., 1994, ApJ, 422, 443\\
Coles P., 1989, MNRAS, 238, 319\\
Coles P., Melott A.L., Shandarin S.F., 1993, MNRAS, 260, 765\\
Collins C.A., Guzzo L., Nichol R.C., Lumsden S.L., 1995, MNRAS,
274, 1071\\
Croft R.A.C., Efstathiou G., 1994, MNRAS, 267, 390\\
Dalton G.B., Croft R.A.C., Efstathiou G., Sutherland W.J., Maddox
S.J., Davis M., 1994, MNRAS, 271, L47\\
Davis M., Summers F.J., Schlegel M., 1992, Nature, 359, 393\\
Dekel A., West M.J., 1985, ApJ, 288, 411\\
Doroshkevich A.G., Shandarin S.F., 1978, MNRAS, 182, 27\\
Efstathiou G., Bond J.R., White S.D.M., 1992, MNRAS, 258 1p\\
Eke V.R., Cole S., Frenk C.S., 1996a, 282,263\\
Eke V.R., Cole S., Frenk C.S., Navarro J.F. 1996b, 
preprint astro-ph/960291\\
Evrard A.E., Metzler C.A., Navarro J.F., 1996, ApJ, 469, 494\\
Fang L.Z., Li S.X., Xiang S.P., 1984, A\&A, 140, 77\\
Gazta\~naga E., Croft R.A.C., Dalton G.B., 1995, MNRAS,
276, 336\\
Gazta\~naga E., Yokoyama J., 1993, ApJ, 403, 450\\
Ghigna S., et al., 1994, ApJ, 437, L71\\
G\'orski K.M., Banday A.J., Bennett C.L., Hinshaw W.G.,
Kogut A., Smoot G.F., Wright E.L., 1996, ApJ, 464, L11\\
G\'orski K.M., Ratra B., Sugiyama N., Banday A., 1995, ApJ, 444,
L65\\
Hockney R.W., Eastwood J.W., 1981, Computer Simulations Using
Particles. McGraw--Hill, New York\\
Hodges H.M., Blumenthal G.R., Kofman L.A., Primack J.R., 
1990, Nucl. Phys. B, 335, 197\\
Holtzman J.A., 1989, ApJS, 71, 1\\
Holtzman J.A., Primack J.R., 1993, ApJ, 405, 428\\
Jing Y.P., B\"orner G., Valdarnini R., 1995, MNRAS, 277, 630\\
Jing Y.P., Fang L.Z., 1994, ApJ, 432, 438\\
Jing Y.P., Plionis M., Valdarnini R., 1992, ApJ, 389, 499\\
Katgert P., et al., 1996, A\&A, 310, 8\\
Kennicutt R.C., Freedman W.L., Mould J.R., 1995, AJ, 110, 1476\\
Kerscher M., Schmalzing J., Retzlaff J., Borgani S., Buchert T., 
Gottl\"ober S., M\"uller V., Plionis M., Wagner H., 1996, MNRAS, in press, 
preprint astro-ph/9606133\\
Klypin A., Borgani S., Holtzman J., Primack J., 1995, ApJ, 444, 1\\
Klypin A., Holtzman J., Primack J., Reg\"os E., 1993, ApJ,
416, 1\\
Klypin A.A., Kopilov A.I., 1983, Sov. Astr. Lett., 9, 41\\
Klypin A., Primack J.R., Holtzman J., 1996, ApJ, 466, 13\\
Klypin A., Rhee G., 1994, ApJ, 428, 399\\
Kofman L., Bertschinger E., Gelb J.M., Nusser A., Dekel A.,
1994, ApJ, 420, 44\\
Kolatt T., Dekel A., Primack J.R., 1996, to be submitted\\
Lacey C., Cole S., 1993, MNRAS, 262, 627\\
Lacey C., Cole S., 1994, MNRAS, 271, 676\\
Liddle A.R., Lyth D.H., 1993, Phys. Rep., 231, 1\\
Liddle A.R., Lyth D.H., Schaefer R.K., Shafi Q., Viana P.T.P., 
1995a, preprint astro-ph/9511057\\
Liddle A.R., Lyth D.H., Viana P.T.P., White M., 1995b, preprint
astro-ph/9512102\\
Lilje P.B., 1992, ApJ, 386, L33\\
Ling E.N., Frenk C.S., Barrow J.D., 1986, MNRAS, 23, 21p\\
Loveday J., Efstathiou G., Peterson B.A., Maddox S.J., 1992,
ApJ, 400, L43\\
Lucchin F., Matarrese S., 1985, Phys. Rev., D32, 1316\\
Lumsden S.L., Heavens A.F., Peacock J.A., 1989, MNRAS, 238, 293\\
Ma C.-P., Bertschinger E., 1994, ApJ 434, L5\\
Maddox S.J., Efstathiou G., Sutherland W.J., Loveday J., 1990,
MNRAS, 242, 43p\\
Mann R.G., Heavens A.F., Peacock J.A., 1993, MNRAS, 263, 798\\
Matarrese S., Lucchin F., Messina A., Moscardini L., 1991,
MNRAS, 252, 35\\
Melott A.L., Pellman T.F., Shandarin S.F., 1994, MNRAS, 269, 626\\
Mo H.J., Jing Y.P., White S.D.M., 1996, preprint astro-ph/9602052\\
Monaco P., 1995, ApJ, 447, 23\\
Moscardini L., Branchini E., Tini Brunozzi P., Borgani S.,
Plionis M., Coles P., 1996, MNRAS, 282, 384\\
Moscardini L., Tormen G., Matarrese S., Lucchin F., 1995, ApJ,
442, 469\\
Mould J., et al., 1995, ApJ, 449, 413\\
Nichol R.C., Briel U.G., Henry J.P., 1994, MNRAS, 265, 867\\
Nichol R.C., Collins C.A., Guzzo L., Lumsden S.L., 1992,
MNRAS, 255, p21\\
Olivier S.S., Primack J.R., Blumenthal G.R., Dekel A., 1993,
ApJ, 408, 17\\
Park C., Vogeley M.S., Geller M.J., Huchra J.P., 1994, ApJ,
431, 569\\
Peacock J.A., Dodds S.J., 1994, MNRAS, 267, 1020\\
Peacock J.A., West M.J., 1992, MNRAS, 259, 494\\
Peebles P.J.E., 1980, The Large Scale Structure of the
Universe. Princeton University Press, Princeton\\
Peebles P.J.E., 1993, Principles of Physical Cosmology.
Princeton University Press, Princeton\\
Plionis M., Borgani S., Moscardini L., Coles P., 1995, ApJ,
441, L57\\
Plionis M., Valdarnini R., 1991, MNRAS, 249, 46 (PV)\\
Plionis M., Valdarnini R., 1995, MNRAS, 272, 869\\
Plionis M., Valdarnini R., Jing Y.P., 1992, ApJ, 398, 12 (PVJ)\\
Pogosyan D.Yu., Starobinsky A.A., 1995, ApJ, 447, 465\\
Postman M., Huchra J.P., Geller M., 1992, ApJ, 384, 407\\
Postman M., Spergel D.N., Satin B., Juszkiewicz R., 1989,
ApJ, 346, 588\\
Press W.H., Schechter P., 1974, ApJ, 187, 452 (PS)\\
Primack J.R., Holtzman J., Klypin A., Caldwell D.O., 1995, Phys.
Rev. Lett., 74, 2160\\
Quintana H., Ramirez A., 1995, ApJSS, 96, 343\\
Raychaudhury S., 1989, Nature, 342, 251\\
Romer A.K., Collins C.A., B\"ohringer H., Cruddace R.C.,
Ebeling H., MacGillawray H.T., Voges W., 1994, Nature, 372, 75\\
Ross G.G., Sarkar S., 1996, Nucl. Phys., B461, 597\\
Sahni V., Coles P., 1995, Phys. Rep., 262, 2\\
Sandage, A., et al., 1996, ApJ, 460, L15\\
Sathyaprakash B.S., Sahni V., Munshi D., Pogosyan D., Melott A.L.,
1995, MNRAS, 275, 463\\
Scaramella R., Baiesi--Pillastrini G., Chincarini G.,
Vettolani G., Zamorani G., 1989, Nature, 338, 562\\
Scaramella R., Zamorani G.,  Vettolani G., Chincarini G., 1990,
AJ, 101, 342\\
Shandarin S.F., Zel'dovich Ya.B., 1989, Rev. Mod. Phys., 61,
185\\
Shapley H., 1930, Harvard Obs. Bull., 874, 9\\
Smoot G.F., et al., 1992, ApJ, 396, L1\\
Stompor R., 1994, A\&A, 287, 693
Stompor R., G\'orski K.M., Banday A.J., 1995, MNRAS, 277, 1225\\
Sugiyama N., 1995, ApJS, 100, 281; ApJ, 444, 489\\
Sutherland W., 1988, MNRAS, 234, 159\\
Tini Brunozzi P., Borgani S., Plionis M., Moscardini L.,
Coles P., 1995, MNRAS, 277, 1210\\
Tormen G., Moscardini L., Lucchin F., Matarrese S., 1993, ApJ,
411, 16\\
Turner M.S., White M., Lidsey J.E., 1993, Phys. Rev. D, 48, 4613\\
van Dalen A., Schaefer R.K., 1992, ApJ, 398, 33\\
Viana P.T.P., Liddle A.R., 1995, preprint astro-ph/9511007\\
Walter, C., Klypin A., 1996, ApJ, 462, 13\\
West M.J., 1989, ApJ, 344, 535\\
White M., Scott D., Silk J., Davis M., 1995, MNRAS, 276, L69\\
White S.D.M., Efstathiou G., Frenk C.S., 1993, MNRAS, 262, 1023\\
White S.D.M., Frenk C.S., Davis M., Efstathiou G., 1987, ApJ,
313, 505\\
Wright E.L., et al., 1992, ApJ, 396, L13\\
Zel'dovich Ya. B., 1970, A\&A, 5, 84\\


\end{document}